\documentclass[10pt]{iopart}

\usepackage{mathrsfs}
\usepackage{iopams}
\usepackage{cite}
\usepackage{epsfig}
\usepackage{graphicx}
\usepackage{color}

\def\3nab{\tilde{\nabla}}
\def\sq{\square}

\def\hsp5{\hspace{5mm}}
\newcommand{\sfrac}[2]{{\textstyle{#1\over#2}}}
\def\case#1/#2{\textstyle\frac{#1}{#2}}

\def\be {\begin{equation}}
\def\ee {\end{equation}}
\def\bea {\begin{eqnarray}}
\def\eea {\end{eqnarray}}

\def\case#1/#2{\textstyle\frac{#1}{#2} }
\def\rf#1{(\ref{#1})}
\def\cqg{{\em Class. Quantum Grav.\/} }

\def\prd{{\em Phys. Rev.\/} {\bf D}}
\def\prl{{\em Phys. Rev. Lett.\/} }
\def\plb{{\em Phys. Lett.\/} {\bf B}}

\def\aph{{\em Ann. Phys. (NY)\/} }

\begin{document}

\title{Cosmological dynamics of Scalar--Tensor  Gravity}

\author[ S Carloni  {\it et al.}]{  S Carloni  \dag, J A Leach \dag, S Capozziello \ddag,
P K S Dunsby \dag \S} 

\address{\dag \ Department of Mathematics and Applied Mathematics,
University of Cape Town, Rondebosch,
7701, South Africa}

\address{\ddag \ Dipartimento di Scienze Fisiche and INFN Sez. di Napoli, Universit\`{a} di Napoli `Federico
II', Complesso Universitario di Monte S Angelo, Via Cinthia I-80126
Napoli, Italy}

\address{\S \  South African Astronomical Observatory, Observatory, Cape Town, South Africa}

\date{\today}

\eads{sante.carloni@gmail.com, capozziello@na.infn.it,
leachj@maths.uct.ac.za and pksd@maths.uct.ac.za}

\begin{abstract}
We study the phase--space of FLRW models derived from Scalar--Tensor
Gravity where the non--minimal coupling is $F(\phi)=\xi\phi^2$  and
the effective potential is $V(\phi)=\lambda \phi^n$. Our analysis
allows to unfold many feature of the cosmology of this class of
theories. For example, the evolution mechanism towards  states
indistinguishable from GR is recovered and proved to depend
critically on the form of the potential $V(\phi)$. Also, transient
almost--Friedmann phases evolving towards accelerated expansion and
unstable inflationary phases evolving towards stable ones are found.
Some of our results are shown to hold also for the String-Dilaton
action.
\end{abstract}

\pacs{98.80.JK, 04.50.+h, 05.45.-a}

\section{Introduction}

Scalar--Tensor Gravity (STG) has become a major area of activity in
the last thirty years. This is due mainly to the discovery of
several shortcomings of General Relativity (GR) in cosmology and
quantum field theory. In fact, the presence of a big bang
singularity, the flatness and horizon problems, led to the
realization that the ``standard" model for cosmology
\cite{weinberg}, based on GR and the standard model of particle
physics, is not able to describe the universe in extreme regimes.
On the other hand, GR is a {\it classical} theory which has so far
eluded every attempt to be quantized, making the
description of gravity on microscopic scales very difficult. This has led many
authors to believe that the standard Einstein scheme is inadequate
to fully describe the gravitational interaction at a fundamental
level \cite{STGrav} and has stimulated the proposal and the study of
alternative theories of gravity that might be able to solve these
problems and at the same time reduce to GR in the weak field
regime.

One of the most fruitful theories of this type is STG, in which
scalar field(s) that are non--minimally coupled to gravity are
introduced. These type of theories were proposed nearly half a
century ago by Jordan \cite{Jordan} and later refined by Brans and
Dicke \cite{BD61}. The original motivation behind Brans-Dicke theory
(BD) came from Mach's principle, but over the years, BD and STG have
gained interest in a wide number of scenarios. For example, in
unification schemes such as superstrings, supergravity or grand
unified theories, the one--loop or higher--loop corrections in the
high--curvature regime take the form of non--minimal couplings to
the geometry or higher--order curvature invariants in their low
energy effective actions \cite{buchbinder}.

In cosmology, STG acquired considerable interest because they
introduce naturally scalar fields and scalar fields which are capable, among other things, of
giving rise to inflationary behaviour \cite{guth}. In GR
the introduction of this type of field has the drawback of raising
the issue of explaining its origin. Instead, in STG this problem
finds a natural solution, because the scalar field can be considered
an additional degree of freedom of the gravitational interaction.
For these reasons, among others, inflationary models based on STG
have been widely studied \cite{La89,La89a}.

Recently, STG has also been used to model dark energy, because
scalar fields are also the natural candidates for phantom and
quintessence fields (see \cite{quintessence} for a review). This
suggests that both inflation and dark energy could be the result of
the action of the same scalar field. STG offers, in this sense, the
ideal framework to implement this intriguing idea and might even
allow us to overcome the problems due to the energy scale difference
between the dark energy scalar fields and the inflaton
\cite{inflessence}.

In this paper, we consider the problem of determining the global
dynamics of Friedmann--Lema\^ {i}tre--Robertson--Walker (FLRW)
cosmologies of STG. We will study a generic class of STG theories,
where quadratic non--minimal couplings to gravity and
self--interaction power--law potentials are assumed. This class of
models is strictly related to the String Dilaton action and
naturally exhibit duality in the cosmological solutions
\cite{dualcap}. Furthermore, they can be obtained from generic
non--minimally coupled scalar--tensor Lagrangians if Noether
symmetries are found in the dynamics \cite{noether,noether1}.
Several exact solutions of these models have been found, but the
stability and global behaviour is still not well understood. Our aim
is to give a full description of the global dynamics of this class
of STG and determine if cosmic histories are possible that (i)
present a transient matter--dominated Friedmann phase and then
evolve towards an accelerated (dark energy--dominated or
$\Lambda$CDM) regime or (ii) present  a first unstable inflationary
phase and a second inflationary attractor.

The dynamical systems approach has been already used with great
success in BD \cite{Oliveira,Kolitch95,Kolitch96,Santos97,Holden98},
in specific models of STG
\cite{Foster98,Amendola90,Billyard99,Gunzig00,Gunzig01,Saa01} and in
other contexts \cite{Carloni05,Leach06a,Laur} allowing the
derivation of new exact solutions and making possible a qualitative
global description of the dynamics.  The geometric structure and
dimensionality of the phase--space of general couplings and
potentials has recently been investigated by Faraoni
\cite{Faraoni05}.

The outline of this paper is as follows. In section 2, we derive the
field equations and discuss the non--minimal coupling and the
self--interacting potential.  The cosmological equations for a FLRW
metric are derived in section 3. We convert them into an autonomous
dynamical system in order to pursue a phase--space investigation.
Sections 4 and 5 are respectively devoted to the analysis of this
system in the absence and presence of a perfect fluid source. We
study the stability of the fixed points with the aim of identifying
physically interesting cosmological behaviours. In section 6, the
results are discussed and some conclusions are drawn.

The following conventions are adopted: the metric signature is $(+ -
- -)$; Latin indices run from $0$ to $3$; $\nabla$  is the covariant
derivative; physical units $c=8\pi G=1$ are used.

\section{The field equations }

In four dimensions, a general action in which gravity is
non--minimally coupled to a scalar field $\phi$, reads
\cite{noether1}:
\begin{equation}
{\cal A}=\int dx^4 \sqrt{-g} \left[ F(\phi)R
+\sfrac{1}{2}g^{ab}\nabla_a\phi\; \nabla_b\phi-V(\phi)+{\cal
L}_M\right], \label{action:F(phi)}
\end{equation}
where $F(\phi)$ is a generic coupling, $V(\phi)$ is the
self--interaction potential and ${\cal L}_M$ is the matter
Lagrangian.

By varying the action
(\ref{action:F(phi)}) with respect to the metric $g_{ab}$, we obtain the field equations
\begin{eqnarray}
&& F(\phi)R_{ab} -\sfrac{1}{2}F(\phi)R
g_{ab}+\sfrac{1}{2}\nabla_a\phi\;
\nabla_b\phi-\sfrac{1}{4}g_{ab}\nabla^c\phi\;
\nabla_c\phi  \nonumber \\
&& +\sfrac{1}{2}g_{ab} V(\phi)-\nabla_a\nabla_b F(\phi)+g_{ab} \sq
F(\phi)=-T^M_{ab}  \label{field:F(phi)}
\end{eqnarray}
and the variation with respect to $\phi$ gives the Klein-Gordon
equation
\begin{equation}
\sq \phi - R F'(\phi)+V'(\phi)=0, \label{field(phi):F(phi)}
\end{equation}
where primes denote differentiation with respect to $\phi$. The
field equations \rf{field:F(phi)} can be recast in the standard form
\begin{equation*}
G_{ab}=R_{ab}-\sfrac{1}{2}Rg_{ab}=T^{eff}_{ab},
\end{equation*}
where the  effective stress-energy
momentum tensor $T^{eff}_{ab}$ is given by
\begin{eqnarray}
T^{eff}_{ab}&=&F(\phi)^{-1}\;
\left[-\sfrac{1}{2}\nabla_a\phi\; \nabla_b\phi+\sfrac{1}{4}g_{ab}\nabla^c\phi\;
\nabla_c\phi -\sfrac{1}{2}g_{ab} V(\phi) \right. \nonumber \\
&&+\left. \nabla_a\nabla_b
F(\phi)-g_{ab} \sq F(\phi)
-T^M_{ab}\right]. \label{stress_energy:effective}
\end{eqnarray}
This form of the equations shows clearly one of the most interesting
features of Scalar Tensor Gravity: the Newtonian gravitational
constant $G_N$, intended as the coupling of gravity with standard
matter, has to be replaced by an effective gravitational "constant"
$G_{eff}$ that depends on the non--minimal coupling $F$ (and, as a
consequence, on $\phi$) and {\em varies} in time.

The Bianchi identities $G^a{}_{b;a}=0$ give the conservation laws
for both the matter and the scalar field. As a general result
\cite{noether1}, it is possible to show that the conservation law
for the scalar field is the Klein-Gordon equation.

The action (\ref{action:F(phi)}) is very general and contains
several interesting physical cases. For example, considering the
transformation
\begin{equation}\label{Trasf:String}
\phi \,\rightarrow \,\exp[-\psi],\quad F(\phi) \,\rightarrow
\,\frac{1}{2}\exp[-2\psi],\quad V(\phi) \,\rightarrow
\,\frac{1}{2}\Lambda\exp[-2\psi],
\end{equation}
which specifies a particular form of coupling and potential, leads to the
4D-String-Dilaton Effective Action
\begin{equation}
{\cal A}=\int dx^4 \sqrt{-g} e^{-\psi}\left[R
+g^{ab}\nabla_a\psi\; \nabla_b\psi-\Lambda\right], \label{action:string}
\end{equation}
where $\Lambda$ is the string charge. In this context, such a theory
is nothing else but a particular STG \cite{capstringpla}. This means
that the considerations and results for the action
(\ref{action:F(phi)}), and the related dynamics, also hold for
string-dilaton cosmology. On the other hand, the set of
transformations
\begin{equation}
F(\phi) \,\rightarrow \,\psi,\quad
\frac{F(\phi)}{2F'(\phi)^2},\rightarrow \,-\omega(\psi),\quad
V(\phi) \,\rightarrow \,0,
\end{equation}
give rise to the action
\begin{equation}
{\cal A}=\int dx^4 \sqrt{-g} \left[\psi R
-\frac{\omega(\psi)}{\psi}g^{ab}\nabla_a\psi\; \nabla_b\psi \right],
\label{action:BD}
\end{equation}
which is nothing else but the BD action\footnote{To be precise, the
proper BD action is exactly recovered only for $\omega =$ constant.}
\cite{BD61}. In addition to the ones above other interesting kind of
transformations are possible \cite{Odintsov}.

In this paper, we will focus on models where the non--minimal
coupling has the form $F(\phi)=\xi\phi^2$ and the self-interaction
potential is taken to be an arbitrary power law of the form
$V(\phi)=\lambda \phi^n$. This choice is general and motivated by
several mathematical and physical reasons. In particular, beside the
string-dilaton and BD actions, several effective quantum field
theories, in low energy physics, can be related to such couplings and
self--interacting potentials  \cite{STGrav,birrell}.
Furthermore, this coupling and potential satisfy the requirement of
Noether symmetries for the Lagrangian in (\ref{action:F(phi)}),
giving rise to general exact solutions of physical interest
\cite{noether}.

A final remark concerns the parameters of the theory. Very different
models can be parameterized by the set $(\xi,\lambda,n)$, but not
all the combinations of these parameters are necessarily physical.
For example, attractive gravity  is  achieved for $\xi<0$ and
$\lambda > 0$, although physically interesting situation can also be
achieved for $\xi>0$, $\lambda < 0$ \cite{cqg97}. In this paper, we
will consider only the physical case $\xi<0$, $\lambda > 0$. The
discussion of the other interesting combinations of parameters can
be found in \cite{leachphd}.

\section{The FLRW dynamical system}

In order to analyze the phase-space of the Scalar Tensor FLRW
cosmologies, the field equations need to be recast in a dynamical
system form. In the FLRW metric and with our choice of coupling and
potential the   the above Einstein-Klein-Gordon equations reduce to:
\begin{eqnarray}
\frac{\ddot{a}}{a}+H\frac{\dot{\phi}}{\phi}+\frac{\ddot{\phi}}{\phi}+
\left(1-\frac{1}{6\xi}\right)\frac{\dot{\phi}^2}{\phi^2}+\frac{\lambda}{6\xi}\phi^{n-2}
-\frac{\mu}{6\xi
\phi^2}(1+3w)=0, \label{Friedmann1:phi2} \\
H^{2}+\frac{k}{a^2}+2H\frac{\dot{\phi}}{\phi}+\frac{\lambda}{6\xi}\phi^{n-2}+\frac{1}{12\xi}
\frac{\dot{\phi}^2}{\phi^2}+\frac{\mu}{3\xi \phi^2}=0,
\label{Friedmann2:phi2} \\
\frac{\ddot{\phi}}{\phi}+3H\frac{\dot{\phi}}{\phi}+12\xi\left(\frac{\ddot{a}}{a}+\frac{\dot{a}^2}{a^2}+\frac{k}{a^2}
\right)+n\lambda \phi^{n-2}=0, \label{KleinGordon:phi2}
\end{eqnarray}
where the first is the cosmological equation for the acceleration,
the second is the Hamiltonian constraint, i.e. the $\{0,0\}$
equation, and the third is the Klein-Gordon equation. $H=\dot{a}/a$
is the Hubble parameter and  $k$ is the spatial curvature constant.
We have also assumed standard matter to be a perfect fluid with a
barotropic index $w$, so that the conservation equation yields
\begin{equation}\label{CE:perfect}
\dot{\mu}=-3 H \mu (1+w)\,,
\end{equation}
where $\mu$ is the the  matter-energy density. In what follows,
since we want to stress the role of a non minimally coupled scalar
field in the modelling of dark energy, we will consider only
$0<w<1$. Other values of these parameters, which can be associated
with more exotic forms of matter energy densities,
although interesting, will not be considered here.

The equations above can be converted into an autonomous system of
first-order differential equations by defining the following set of
expansion normalised variables:
\begin{eqnarray}\label{phi2:var}
x = \frac{\dot{\phi}}{\phi H},\ \ \ \ \ \  &&y = \frac{\lambda \phi^{n-2}}{6\xi H^2}, \nonumber \\
z = \frac{\mu}{3\xi \phi^2 H^2}, \ \ \ \ \ \ &&K =\frac{k}{a^2 H^2}.
\end{eqnarray}
from which we obtain
\begin{eqnarray}\label{phi2:eqn_mat}
&&\fl x' =
\frac{1}{12\xi-1}\left[12\xi\left(1+K+Kx\right)+2(1+6\xi)x+(5-24\xi)x^2-\left(1-\sfrac{1}{6\xi}\right)x^3
\right.\nonumber \\
&&\left.+6\xi (n-2)y-(1-6n\xi)xy+6\xi(1+3w)z+\frac{1}{2}(1+3w)xz\right], \nonumber \\
&&\fl y'=\frac{2y}{12\xi-1}\left[24\xi-1+12\xi K +4x+(12\xi-1)(n-2)x\right. \nonumber \\
&&\fl \left.-\left(1-\sfrac{1}{6\xi}\right)x^2 -(1-6n\xi)y+\frac{1}{2}(1+3w)z \right], \\
&&\fl z' = \frac{z}{12\xi-1}\left[7(1-12\xi)-2-3w(1-12\xi)+24\xi K +6(1-4\xi)x \right. \nonumber \\
&&\fl \left. -2\left(1-\sfrac{1}{6\xi}\right)x^2-2(1-6n\xi)y+(1+3w)z\right] , \nonumber \\
&&\fl
K'=\frac{2K}{12\xi-1}\left[12\xi(1+K)+2x-\left(1-\sfrac{1}{6\xi}\right)x^2-(1-6n\xi)y+\frac{1}{2}(1+3w)z\right],
\nonumber
\end{eqnarray}
where primes denote derivatives with respect to a new evolution variable
$\tau=\ln a$ and the dynamical variables are constrained by
\begin{equation}\label{phi2:constraint}
1+2x+\sfrac{1}{12\xi}x^2+y+K-z=0.
\end{equation}
The associated phase-space is 4-dimensional and the evolution is
constrained by (\ref{phi2:constraint}). The task is now to study the
structure of such a space: this means finding the stability of the
fixed points, and then to analyze the evolution of trajectories
\cite{Dynamical}. We will consider two cases: the vacuum case
($\mu=0$) and the matter case ($\mu\neq 0$).

\section{The vacuum case}

When we consider the vacuum case ($\mu=0$), the set of dynamical
equations \rf{phi2:eqn_mat} reduces to
\begin{eqnarray}\label{phi2:eqn_vac}
\hspace{-15 mm} x' = \frac{1}{12\xi-1}
\left[12\xi\left(1+K+Kx\right)+2(1+6\xi)x+(5-24\xi)x^2-\left(1-\sfrac{1}{6\xi}\right)x^3
\right.
\nonumber \\
\left.
+6\xi (n-2)y-(1-6n\xi)xy\right], \nonumber \\
\hspace{-15 mm}y'=\frac{2y}{12\xi-1}\left[24\xi-1+12\xi K +4x+(12\xi-1)(n-2)x\right.  \\
\left.-\left(1-\sfrac{1}{6\xi}\right)x^2 +(6n\xi-1)y\right], \nonumber \\
\hspace{-15 mm}
K'=\frac{2K}{12\xi-1}\left[12\xi(1+K)+2x-\left(1-\sfrac{1}{6\xi}\right)x^2+(6n\xi-1)y\right],
\nonumber
\end{eqnarray}
with the constraint equation  given by
\begin{equation}\label{phi2:constraint_vac}
1+2x+\sfrac{1}{12\xi}x^2+y+K=0\,.
\end{equation}
In this case, the phase-space is 3-dimensional.

\subsection{Finite analysis}

We can further simplify the system \rf{phi2:eqn_vac} by implementing
the constraint \rf{phi2:constraint_vac}:
\begin{eqnarray}\label{phi2:eqn_vac2}
\hspace{-5 mm} x' = -2x-4x^2-\sfrac{1}{6\xi}x^3
+\frac{y}{12\xi-1}\left[6\xi (n-4)+(6\xi(n-2)-1)x\right], \\
\hspace{-5 mm}y'=y\left[2+(n-6)x-\sfrac{1}{3\xi}x^2
+\frac{2(6\xi(n-2)-1)}{12\xi-1}y\right]. \nonumber
\end{eqnarray}
From these equations, it is clear that the $x$ axis ($y=0$),
characterized by the absence of a potential for the scalar field, is
an invariant submanifold. This tells us that there is no orbit for
which the potential of the scalar field can become exactly zero and
that if the potential is initially set to zero, it will remain zero.

The fixed points can be obtained by setting $x'=0$ and $y'=0$. For
the system \rf{phi2:eqn_vac2} we obtain five fixed points (see Table
\ref{T:fixed_vac}). The coordinate of the point $\mathcal{A}$ is
independent of $\xi$ and $n$ and the ones of $\mathcal{B}$ and
$\mathcal{C}$ are independent of $n$. The point $\mathcal{D}$ is a
finite fixed point for $n\neq2$ and $\mathcal{E}$ a finite fixed
point for $\xi\neq1/2(n+2)$.

Merging occurs for the points $\mathcal{E}$ and $\mathcal{B}$, and
$\mathcal{E}$ and $\mathcal{C}$ for $n=4\mp\frac{\sqrt{3} \sqrt{\xi
(12 \xi -1)}}{\xi }$ respectively. The point $\mathcal{E}$ also
merges with $\mathcal{D}$ for $n=4\pm\sqrt{\frac{12 \xi -1}{\xi }}$.

All the fixed points except $\mathcal{A}$ and $\mathcal{D}$ are
associated with flat spatial geometry. Point $\mathcal{A}$ is
associated with an open spatial geometry and for $\mathcal{D}$ the
sign of the space curvature depends on $\xi$ and $n$: $K_{
\mathcal{D}}$ is positive for $n\neq2$, $4-\sqrt{\frac{12 \xi
-1}{\xi }}<n<4+\sqrt{\frac{12 \xi -1}{\xi }}$ and negative
otherwise.

The stability of the fixed points can be determined by evaluating
the eigenvalues of the Jacobian matrix  associated with the system
\rf{phi2:eqn_vac2} (see Table~\ref{T:eigenvalues_vac}), as
prescribed by the Hartman-Grobman Theorem \cite{Hartman}
\footnote{The values of the parameter for which the  eigenvalues are
zero are bifurcations for the dynamical system. In this paper we
will not give an analysis of the bifurcations referring the reader
to the specific literature for more details.}.

The fixed point $\mathcal{A}$ is a saddle for every value of the
parameters $\xi$ and $n$. The point $\mathcal{B}$ can either be a
stable node or a saddle node whereas $\mathcal{C}$ is either an
unstable node or a saddle node, depending on the values of $\xi$ and
$n$. The eigenvalues of $\mathcal{D}$ and $\mathcal{E}$ are both
dependent on $\xi$ and $n$ so that the stability varies over the
different ranges of $\xi$ and $n$. The stability of these fixed
points have been summarized in Table~\ref{T:stability_vac}.

The coordinates of the fixed points may be used to determine exact
cosmological solutions at the fixed points themselves. In fact, when evaluated at
these points,  the Friedmann and the Klein-Gordon equations can be
written as:
\begin{eqnarray}
\dot{H}
= -\frac{H^2}{\alpha},\qquad
\alpha=\left[1-2x_i-\sfrac{1}{6\xi}x_i^2+\left(\sfrac{6\xi(
n-2)-1}{12\xi-1}\right)y_i\right]^{-1}, \label{phi2:vac_H}\\
\frac{\ddot{\phi}}{\phi}
=- \beta\ H^2,\qquad \beta=3x_i+x_i^2+\sfrac{6\xi(n-4)}{1-12\xi}y_i
,\label{phi2:vac_phi}
\end{eqnarray}
where $(x_i,y_i)$ represent the coordinates of the fixed points.
Integrating \rf{phi2:vac_H} and substituting into \rf{phi2:vac_phi}
gives
\begin{eqnarray}
\frac{\ddot{\phi}}{\phi}+\frac{\alpha^{2}\beta}{(t-t_{0})^{2}
}=0,\label{phi2:vac_phi1}
\end{eqnarray}
which has a Cauchy-Euler form. If the terms $\alpha$ and $\beta$ are
different from zero, equations \rf{phi2:vac_H} and
\rf{phi2:vac_phi1} can be easily integrated, giving
\begin{eqnarray}
a=a_0\left(t-t_0\right)^\alpha, \label{phi2:vac_sol_gen}\\\nonumber\\
\fl\phi =\left\{ \begin{array}{cc}
      \left(t-t_0\right)^{1/2}\left[\phi_0\left(t-t_0\right)^{m}+\phi_1\left(t-t_0\right)^{-m}\right]\;,,
      & \mbox{if}\;\; \alpha^{2}\beta<\frac{1}{4} \;,\\
      \left(t-t_0\right)^{1/2}\left(\phi_0+\phi_1\ln t\right)\;,
      & \mbox{if}\;\; \alpha^{2}\beta=\frac{1}{4} \;,\\
       \left(t-t_0\right)^{1/2}\left(\phi_0 \sin\left[ m\ln \left(t-t_0\right)\right]+
       \phi_1\cos\left[ m\ln \left(t-t_0\right)\right]\right)\;,
       & \mbox{if}\;\; \alpha^{2}\beta>\frac{1}{4}\;,
    \end{array}
    \right.\label{sol phi}
\end{eqnarray}
where
\begin{equation}\label{phi2:vac_alpha}
m=\frac{1}{2}\sqrt{1-4\alpha^{2} \beta}.
\end{equation}

In the case of  point $\mathcal{A}$,  we have a Milne evolution and
a constant scalar field
\begin{equation}\label{phi2:vac_sol_A}
\alpha_{\mathcal{A}}=1, \qquad \phi_{\mathcal{A}}=\phi_0.
\end{equation}
The above solution (as well as some others that will follow) is
particularly interesting because the scalar field is {\em constant}. This
implies that $\mathcal{A}$ represents a state in which $G_{eff}$ is
constant and the potential of the scalar field acts as a
cosmological constant. In other words, at this point, scalar tensor
gravity is indistinguishable from standard GR plus cosmological
constant.

For the point $\mathcal{B}$ the solutions are given by
\rf{phi2:vac_sol_gen} and \rf{sol phi} with
\begin{equation}\label{phi2:vac_sol_B}
\fl \alpha_{\mathcal{B}}=\frac{1}{3(1-8\xi)-4\sqrt{3\xi(12\xi-1)}}\;
,\qquad m_{\mathcal{B}}=\frac{3}{6(1-8\xi)-8\sqrt{3\xi(12\xi-1)}}\;,
\end{equation}
and for point $\mathcal{C}$
\begin{equation}\label{phi2:vac_sol_C}
\fl \alpha_{\mathcal{C}}=\frac{1}{3(1-8\xi)+4\sqrt{3\xi(12\xi-1)}}\;
,\qquad m_{\mathcal{C}}=\frac{3}{6(1-8\xi)+8\sqrt{3\xi(12\xi-1)}}\;
.
\end{equation}
Note that, for $\xi<0$, the value of $\alpha$ for the above two
solutions is always positive and less than 1 i.e. these two solution
always represent two Friedmann-like solutions. In addition,
$\mathcal{B}$ represents a solution in which the scalar field is
growing, while at $\mathcal{C}$, $\phi$ is dissipating. If $n\neq2$,
the point $\mathcal{D}$ is associated with
\begin{equation}\label{phi2:vac_sol_D}
\alpha_{\mathcal{D}}=1,\qquad m_{\mathcal{D}}=\frac{n+2}{2(n-2)},
\end{equation}
which represents another Milne solution, while the scalar field is
decreasing for $n>2$ and increasing for $n>2$. It is interesting
that, unlike the Milne solution in GR, this linear solution
for $a$ is not necessarily a spatially hyperbolic one. The constant
$a_0$ can be related to the parameters $\xi$ and $n$ for non-flat
solutions:
\begin{equation}\label{phi2:vac_a0_D}
a_0^2=-\frac{k(n-2)^2\xi}{1+[4+n(n-8)]\xi}.
\end{equation}
When $n=2$, $\mathcal{D}$ becomes an asymptotic fixed point and
merges with  $\mathcal{D}_\infty$ (its solution will be presented
within the asymptotic analysis). Finally, for point $\mathcal{E}$ we
have
\begin{equation}\label{phi2:vac_sol_E}
\alpha_{\mathcal{E}}=\frac{2(n+2)\xi-1}{(n-4)(n-2)\xi},\qquad m_{\mathcal{E}}=\frac{n+2}{2(n-2)}
\end{equation}
for all $n\neq2,\;4$. This solution represent an expansion for
$\frac{1-4 \xi }{2 \xi }<n<2$ and $n>4$ and was already found in
other contexts
\cite{Amendola90,Holden:1999hm,Amendola:1999qq,Barrow:1990nv,Gannouji:2006jm}.
In particular, in \cite{Gannouji:2006jm} it is shown that for
$2<n<4$ the scale factor associated to this point evolves towards a
superinflating state (also called ``Big Rip" singularity) without
including any exotic feature like ghosts or non standard fluids.
When $n=2$ and $n=4$, using the cosmological equations we obtain the
solutions
\begin{eqnarray}
a=a_0 \; ,\qquad && \phi=\phi_0\  (\lambda=0)\; , \label{phi2:vac_sol_E2}\\
a=a_0 e^{C(t-t_0)} ,\qquad && \phi=\phi_0= \pm\sqrt{ \sfrac{-6\xi
C^2}{\lambda}}\, , \label{phi2:vac_sol_E4}
\end{eqnarray}
which represent a static universe and a de Sitter evolution
respectively. In both these cases  an effective cosmological constant
is present, whose value depends on  the effective gravitational
constant (via $\phi_0$) and the coupling constant $\lambda$ of the
self-interaction of the scalar field. Again, since the scalar field is
constant in these cases, these solution are indistinguishable from
the GR solutions. The difference with point $\mathcal{A}$ is
that the solution does not occur in ``pure" GR and can
also be stable. This is particularly interesting in the second case
($n=4$) in which a de-Sitter solution able to mimic an inflationary
or dark energy phase in a $\Lambda$GR cosmology is a semi-global
attractor. The possibility that scalar tensor gravity could converge
to GR has been proposed within the context of extended inflation
first using  numerical techniques
\cite{Garcia-Bellido:1990jz} and than with a more formal proof
\cite{Damour:1993id}. The dynamical system approach allows one to see
this phenomenon in a very clear way, even in the more general case of
a non-zero potential. It turns out that the nature of the potential,
determined in our case by the value of $n$, together with the value
of the coupling plays a critical role in the realization of this
mechanism.
\begin{table}
\caption{The coordinates and scale factor solutions of the fixed
points for the vacuum case. We only show the exponent $\alpha$ of
the solutions \rf{phi2:vac_sol_gen}.}
\begin{tabular}{llll}
\br Point &  Coordinates $(x,y)$ & $K$ & $\alpha$
\\ \hline
$\mathcal{A}$ & $\left(0,\; 0\right)$ & $-1$ &  $1$ \\
$\mathcal{B}$ & $\left(2(-6\xi-\sqrt{3\xi(12\xi-1)}),\; 0\right)$
& $0$ & $\frac{1}{3(1-8\xi)-4\sqrt{3\xi(12\xi-1)}}$ \\
$\mathcal{C}$ & $\left(2(-6\xi+\sqrt{3\xi(12\xi-1)}),\; 0\right)$ &
$0$ & $\frac{1}{3(1-8\xi)+4\sqrt{3\xi(12\xi-1)}}$
 \\
$\mathcal{D}$ & $\left(-\frac{2}{n-2},\;
\frac{2-24\xi}{3\xi(n-2)^2}\right)$ &
$\frac{-1-[4+n(n-8)]\xi}{(n-2)^2\xi}$ & $1$,
$\left(a_0^2=-\frac{k(n-2)^2\xi}{1+[4+n(n-8)]\xi}\right)$
 \\
$\mathcal{E}$ & $\left(\frac{2(n-4)\xi}{1-2(n+2)\xi},\;
\frac{(12\xi-1)\left[3+(n-10)(n+2)\xi\right]
}{3(1-2(n+2)\xi)^2}\right)$ & $0$ & $\left\{ \begin{tabular}{ll}
$\frac{2(n+2)\xi-1}{(n-4)(n-2)\xi}$, & $n\neq 2,\;4$ \\
$a=a_0$, & $n= 2$ \\
 $a=a_0 e^{C(t-t_0)}$, & $n=4$ \end{tabular} \right.$ \\
\br
\end{tabular}\label{T:fixed_vac}
\end{table}

\begin{table}[tbp] \centering
\caption{Values of the parameter $m$ and the corresponding scalar
field solutions for the fixed points in vacuum.}
\begin{tabular}{cll}
\br Point & $m$  & Solutions \\ \mr
$\mathcal{A}$ & $\frac{1}{2}$ & $\phi=\phi_1$,\hspace{21mm} $\phi_0,\,\lambda=0$ \\
$\mathcal{B}$ & $\sfrac{3}{6(1-8\xi)-8\sqrt{3\xi(12\xi-1)}}$ & $\phi=\phi_0 (t-t_0)^{1/2+m}$, \ $\phi_1,\, \lambda=0$\\
$\mathcal{C}$ &  $\frac{3}{6(1-8\xi)+8\sqrt{3\xi(12\xi-1)}}$ & $\phi=\phi_1 (t-t_0)^{1/2-m}$, \ $\phi_0,\, \lambda=0$\\
$\mathcal{D}$ & $\frac{n+2}{2(n-2)}$  & $\left\{\begin{tabular}{ll} $\phi=\phi_0^\mathcal{D} (t-t_0)^{1/2}$, & $n=-2$\\
$\phi=\phi_1^\mathcal{D} (t-t_0)^{1/2-m}$, & $n\neq -2,\;2$ \\
$\phi_0^\mathcal{D}=\pm\left(\sfrac{4\lambda}{1-12\xi}\right)^{\frac{1}{4}}$,&
$\phi_1^\mathcal{D}=\left(\sfrac{4(1-12\xi)}{\lambda(n-2)^2}\right)^{\frac{1}{n-2}}$
\end{tabular} \right.$ \\& & \\
$\mathcal{E}$ & $\frac{n+2}{2(n-2)}$  & $\left\{\begin{tabular}{l} $\phi=\phi_0^\mathcal{E} (t-t_0)^{1/2}$,\ \ \ \ $n=-2$\\
$\phi=\phi_1^\mathcal{E} (t-t_0)^{1/2-m}$,\ \ \ \  $n\neq -2,\;2,\;4$  \\
$\phi_0^\mathcal{E}=\pm2\left(\sfrac{6\lambda\xi}{12\xi-1}\right)^{\frac{1}{4}}$,
\\
$\phi_1^\mathcal{E}=\left(\sfrac{2(12\xi-1)[3+(n-10)n+2)\xi]}{\lambda \xi (n-2)(n-4)}\right)^{\frac{1}{n-2}}$ \end{tabular} \right.$\\
& & \\ \br
\end{tabular}\label{T:sol_vac}
\end{table}

\begin{table}[tbp] \centering
\caption{The eigenvalues associated with the fixed points in the
vacuum model.}
\begin{tabular}{ll}
\br Point &  Eigenvalues \\ \hline
& \\
$\mathcal{A}$ &
$\left[-2,2\right]$  \\
$\mathcal{B}$ &
$\left[4(1-12\xi)-8\sqrt{3\xi(12\xi-1)},\; 6-2(n+2)\left(6\xi+\sqrt{3\xi(12\xi-1)}\right) \right]$  \\
$\mathcal{C}$ &
$\left[4(1-12\xi)+8\sqrt{3\xi(12\xi-1)},\; 6-2(n+2)\left(6\xi-\sqrt{3\xi(12\xi-1)}\right) \right]$  \\
$\mathcal{D}$ & $\left[\frac{(4-n)\xi-\sqrt{\xi
\{3(8-n)n\xi-4\}}}{(n-2)\xi},\;
\frac{(4-n)\xi+\sqrt{\xi\{3(8-n)n\xi-4\}}}{(n-2)\xi} \right]$  \\
$\mathcal{E}$ & $\left[\frac{3+(n-10)(n+2)\xi}{2(n+2)\xi-1},\;
\frac{2+2\{4+(n-8)n)\}\xi}{2(n+2)\xi-1} \right]$  \\
& \\ \br
\end{tabular}\label{T:eigenvalues_vac}
\end{table}

\begin{table}[tbp] \centering
\caption{Stability of the fixed points in the vacuum case. The
parameters are $N_{\pm}=4\pm\sqrt{3(12\xi-1)/\xi}$, $P_{\pm}=4\pm
2\sqrt{(12\xi-1)/3\xi}$ and $Q_{\pm}=4\pm \sqrt{(12\xi-1)/\xi}$. We
use the term `attractor' to denote a sink, `repeller' to denote a
source and `spiral' to denote an attractive spiral.}
\begin{tabular}{lcccc}
\br {\small Points}  & {\small $n< \frac{1-4\xi}{2\xi}$} & {\small
$\frac{1-4\xi}{2\xi}\leq n<N_- $}
& {\small $N_{-}<n<P_{-} $} & {\small $P_{-} <n<Q_{-} $} \\
\mr{\small $\mathcal{A}$} & {\small saddle} & {\small
saddle} & {\small saddle} & {\small saddle}\\
{\small $\mathcal{B}$} & {\small repeller} & {\small repeller} &
{\small repeller} & {\small repeller}\\
{\small $\mathcal{C}$} & {\small saddle} & {\small saddle} & {\small
repeller} & {\small
repeller} \\
{\small $\mathcal{D}$} & {\small spiral} & {\small spiral} & {\small spiral} & {\small attractor} \\
{\small $\mathcal{E}$} & {\small attractor} & {\small repeller}  & {\small saddle} & {\small saddle}\\
\br & {\small $Q_{-}< n< 2$} & {\small $2<n<Q_{+} $}
& {\small $Q_{+}<n\leq P_{+} $} & {\small $P_{+}<n<N_{+} $} \\
\mr
{\small $\mathcal{A}$} & {\small saddle} & {\small saddle} & {\small saddle} & {\small saddle}\\
{\small $\mathcal{B}$} & {\small repeller} & {\small repeller} &
{\small repeller} &{\small
repeller} \\
{\small $\mathcal{C}$} & {\small repeller} & {\small repeller} &
{\small repeller} & {\small
repeller}\\
{\small $\mathcal{D}$} & {\small saddle}& {\small saddle} & {\small attractor} & {\small spiral} \\
{\small $\mathcal{E}$} & {\small attractor} & {\small attractor}  & {\small saddle} & {\small saddle} \\
\br & {\small $n> N_{+}$} &
&  &  \\
\mr
{\small $\mathcal{A}$} & {\small saddle} &  &  & \\
{\small $\mathcal{B}$} & {\small saddle} & & & \\
{\small $\mathcal{C}$} & {\small
repeller} &  &  & \\
{\small $\mathcal{D}$} & {\small spiral} & & & \\
{\small $\mathcal{E}$} & {\small
repeller} &   &  &\\
\br
\end{tabular}\label{T:stability_vac}
\end{table}
It is useful to define the deceleration parameter $q$ in terms of
the dynamical variables:
\begin{equation}\label{q}
q=-\frac{\dot{H}}{H^2}-1=2x_i+\sfrac{1}{6\xi}x_i^2-\left(\sfrac{1+12\xi-6\xi
n}{1-12\xi}\right)y_i.
\end{equation}
This equation represents a parabola in the phase space that divides
the accelerating $(q<0)$ expansion phases from the decelerating
$(q>0)$ ones. Points $\mathcal{A}$ and $\mathcal{D}$ lie on the
curve \rf{q} as expected by the form of their scale factor
solutions. On the other hand  $\mathcal{B}$ and $\mathcal{C}$  always
lie on the decelerated expansion side of the curve. Instead, for $\mathcal{E}$
we have decelerated expansion for $4-\sqrt{\frac{12 \xi
-1}{\xi }}<n<2$ or $4<n<4+\sqrt{\frac{12 \xi -1}{\xi }}$ and a
decelerated one for $\frac{1-4 \xi }{2 \xi }<n<4-\sqrt{\frac{12 \xi
-1}{\xi }}$ or $n>4+\sqrt{\frac{12 \xi -1}{\xi }}$.

\subsection{Asymptotic analysis}

Since the dynamical system (\ref{phi2:eqn_vac2}) is not compact, it
might admit an asymptotic structure that is relevant for the
global dynamics. In order to analyze the asymptotic features of the
phase space, we use the Poincar\'{e} projection \cite{Amendola90,
Capozziello93}. This method consists of transforming to the polar
coordinates
\begin{equation}
x=\bar{r}\cos \psi, \ \ \ \ \ y=\bar{r}\sin \psi
\end{equation}
and setting $\bar{r}=\frac{\sqrt{r}}{1-r}$.  In this way, the
asymptotic regime is achieved for $r \rightarrow 1$. Using the
Poincar\'{e} projection,  the asymptotic form of the dynamical
equations (\ref{phi2:eqn_vac2}) read
\begin{eqnarray}
r'&&=  \frac{\cos^2 \psi\; \left(\cos 2\psi-3\right)}{12\xi}, \label{phi2:inf_vac_r} \\
\psi ' &&= -\frac{\cos^3 \psi\; \sin \psi}{6\xi(1-r)^2}.
\label{phi2:inf_vac_phi}
\end{eqnarray}
Note that the radial equation does not depend on the radial
coordinate. This means that the fixed points can be obtained by only
considering the equation for $\psi'$. Setting $\psi '=0$ we obtain
the four fixed points listed in Table~\ref{Table_phi2:vac_asymp}.

\begin{figure}
\begin{center}
\includegraphics[clip=true,scale=0.50]{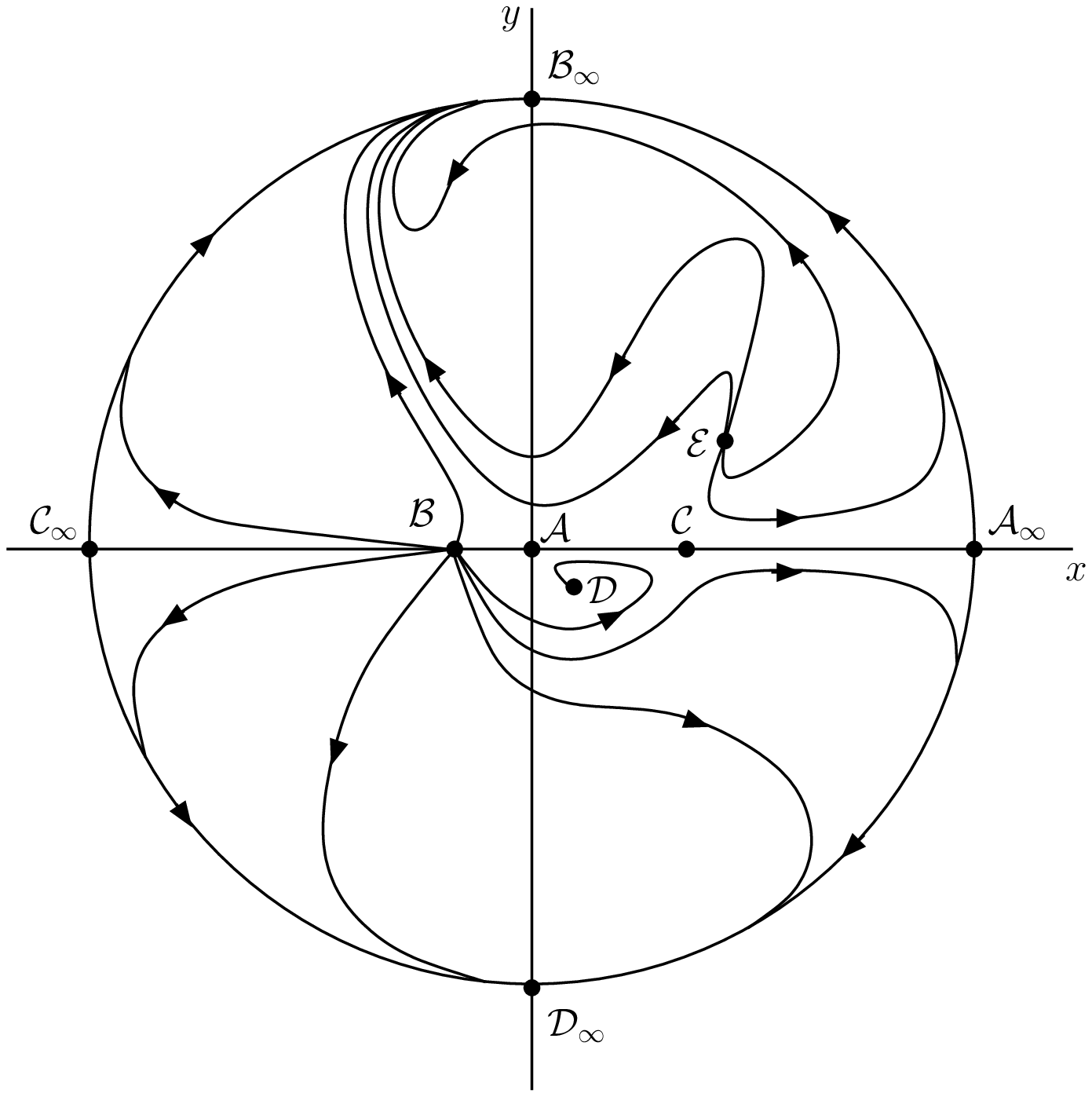} \caption{Global
phase space for
$\frac{1-4\xi}{2\xi}<n<4-\sqrt{3}\sqrt{\frac{12\xi-1}{\xi}}$ (e.g.
$n=-3$ and $\xi=-0.35$). Note that the non hyperbolic point
$\mathcal{B}_\infty$ correspond to a saddle everywhere but on the
border of the circle in which it behaves as an attractor (see the
text for details). }\label{fig1}
\end{center}
\end{figure}

Let us now derive the behaviour of the scale factor at these points.
This can be done by integrating the Friedmann equation
(\ref{phi2:vac_H}) in the above asymptotic limit. The points
$\mathcal{A}_\infty$ and $\mathcal{C}_\infty$ correspond to $y\to 0$
as $x \rightarrow \pm \infty$ respectively. In this limit the $x'$
equation \rf{phi2:eqn_vac2} reduces to
\begin{equation}
x'=-\frac{1}{6\xi}\; x^3\;,
\end{equation}
which admits the solution
\begin{equation}\label{phi2:vac_inf_AC}
x^2=\frac{3\xi}{\tau-\tau_{\infty}}.
\end{equation}
In the given limit, the evolution equation \rf{phi2:vac_H} can be
written as
\begin{equation}
\frac{H'}{H}=\frac{1}{6\xi}\; x^2\;,
\end{equation}
which may be solved together with \rf{phi2:vac_inf_AC} to yield the
solution
\begin{equation}\label{phi2:vac_sol_inf_AC}
|\tau-\tau_{\infty}|=\left[C_1 \pm C_2(t-t_0)\right]^{2}\;,
\end{equation}
which represents a universe which reaches a maximum radius and then
recollapses. The points $\mathcal{B}_\infty$ and
$\mathcal{D}_\infty$, instead, correspond to $x\to 0$ as $y
\rightarrow \pm \infty$. In this limit the second of
\rf{phi2:eqn_vac2} reduces to
\begin{equation}
y'=\frac{2\left(6\xi(n-2)-1\right)}{12\xi-1}\; y^2\;,
\end{equation}
which admits the solution
\begin{equation}\label{phi2:vac_inf_BD}
y=\frac{12\xi-1}{2\left(1-6\xi(n-2)\right)(\tau-\tau_{\infty})}.
\end{equation}
Equation \rf{phi2:vac_H} now reduces to
\begin{equation}
\frac{H'}{H}=\frac{\left(1-6\xi(n-2)\right)}{12\xi-1}y,
\end{equation}
and substituting the solution for $y$, we obtain
\begin{equation}\label{phi2:vac_sol_inf_BD}
|\tau-\tau_{\infty}|=\left[C_1 \pm C_2(t-t_0)\right]^{2}.
\end{equation}
The stability of the asymptotic fixed points is summarized in
Table~\ref{Table_phi2:vac_asymp}. The points $\mathcal{A}_\infty$
and $\mathcal{C}_\infty$ are saddles for every value of $n$ and
$\xi$. The  points $\mathcal{B}_\infty$ and $\mathcal{D}_\infty$ are
non-hyperbolic and they can be shown to represent saddle-nodes. This
means that they behave like saddles or nodes depending on which
direction the orbits approach them and that a local separatrix
exists to divide the different stability domains. In our specific
case this separatrix corresponds to the equator of the Poincar\'{e}
sphere (i.e. our ``infinity"), so that effectively
$\mathcal{B}_\infty$ behaves as a saddle if $n<2+1/6\xi$ or $n=2$
and an attractor if $n>2+1/6\xi$ ($n\neq 2$), and
$\mathcal{D}_\infty$ behaves like a saddle if $n>2+1/6\xi$ (including
$n=2$) and an attractor if $n<2+1/6\xi$. For both these point the
separatrix (i.e. our unitary circle) is always  attractive so that
orbits can bounce off the saddle and then approach the point
along the unitary circle (see for example Figure \ref{fig1}). A
summary of the stability of the asymptotic fixed points is
summarized in Table~\ref{Table_phi2:vac_asymp}.

The behaviour of the scalar field can be obtained in the same way by
substituting the solutions for $x$ and $y$ in the appropriate limit
of equation \rf{phi2:vac_phi}. For the points $\mathcal{A}_\infty$
and $\mathcal{C}_\infty$ \rf{phi2:vac_phi} we obtain
\begin{equation}\label{phi_asyp_AC}
\phi=c_1\cos \left(c_0\sqrt{3\xi}(t-t_0)\right)+c_2\sin
\left(c_0\sqrt{3\xi}(t-t_0)\right),
\end{equation}
where $c_0,c_1$ and $c_2$ are integration constants. For the points
$\mathcal{B}_\infty$ and $\mathcal{D}_\infty$ we obtain
\begin{equation}\label{phi_asyp_BD1}
\fl \phi=c_1\cos
\left(c_0(t-t_0)\sqrt{\left|\sfrac{6\xi(n-4)}{2(1-6\xi(n-4))}\right|}\right)+c_2\sin
\left(c_0(t-t_0)\sqrt{\left|\sfrac{6\xi(n-4)}{2(1-6\xi(n-4))}\right|}\right),
\end{equation}
for $n>2+\sfrac{1}{6\xi}$ or $n>4$, and
\begin{equation}\label{phi_asyp_BD2}
 \phi=c_1 e^
{c_0(t-t_0)\sqrt{\frac{6\xi(n-4)}{2(1-6\xi(n-4))}}}+c_2e^{
c_0(t-t_0)\sqrt{\frac{6\xi(n-4)}{2(1-6\xi(n-4))}}},
\end{equation}
for $2+\sfrac{1}{6\xi}<n<4$.

\begin{table}[tbp] \centering
\caption{Coordinates, behaviour of the scale factor and stability of
the asymptotic fixed points in the vacuum model.}
\begin{tabular}{llll}
\br Point & $\psi$ & Behaviour & Stability
\\ \mr
$\mathcal{A}_\infty$ & $0$ &$|\tau-\tau_\infty|=\left[C_1 \pm
C_2(t-t_0)\right]^{2}$  &
saddle \\
$\mathcal{B}_\infty$ & $\sfrac{\pi}{2}$ &
$|\tau-\tau_\infty|=\left[C_1 \pm C_2(t-t_0)\right]^{2}$
&$\left\{\begin{array}{cc}
\mbox{saddle} & \{n<2+\frac{1}{6\xi}\}+\{n=2\} \\
\mbox{attractor} & \{n>2+\frac{1}{6\xi}\}-\{n=2\}
\end{array}
\right.$ \\
$\mathcal{C}_\infty$ & $\pi$ & $|\tau-\tau_\infty|=\left[C_1 \pm
C_2(t-t_0)\right]^{2}$  & saddle \\
$\mathcal{D}_\infty$ & $\sfrac{3\pi}{2}$ &
$|\tau-\tau_\infty|=\left[C_1 \pm C_2(t-t_0)\right]^{2}$  &
$\left\{\begin{array}{cc}
\mbox{attractor} & n<2+\frac{1}{6\xi} \\
\mbox{saddle} & n>2+\frac{1}{6\xi}
\end{array}
\right.$ \\
\br
\end{tabular}\label{Table_phi2:vac_asymp}
\end{table}

\begin{figure}
\begin{center}
\includegraphics[clip=true,scale=0.50]{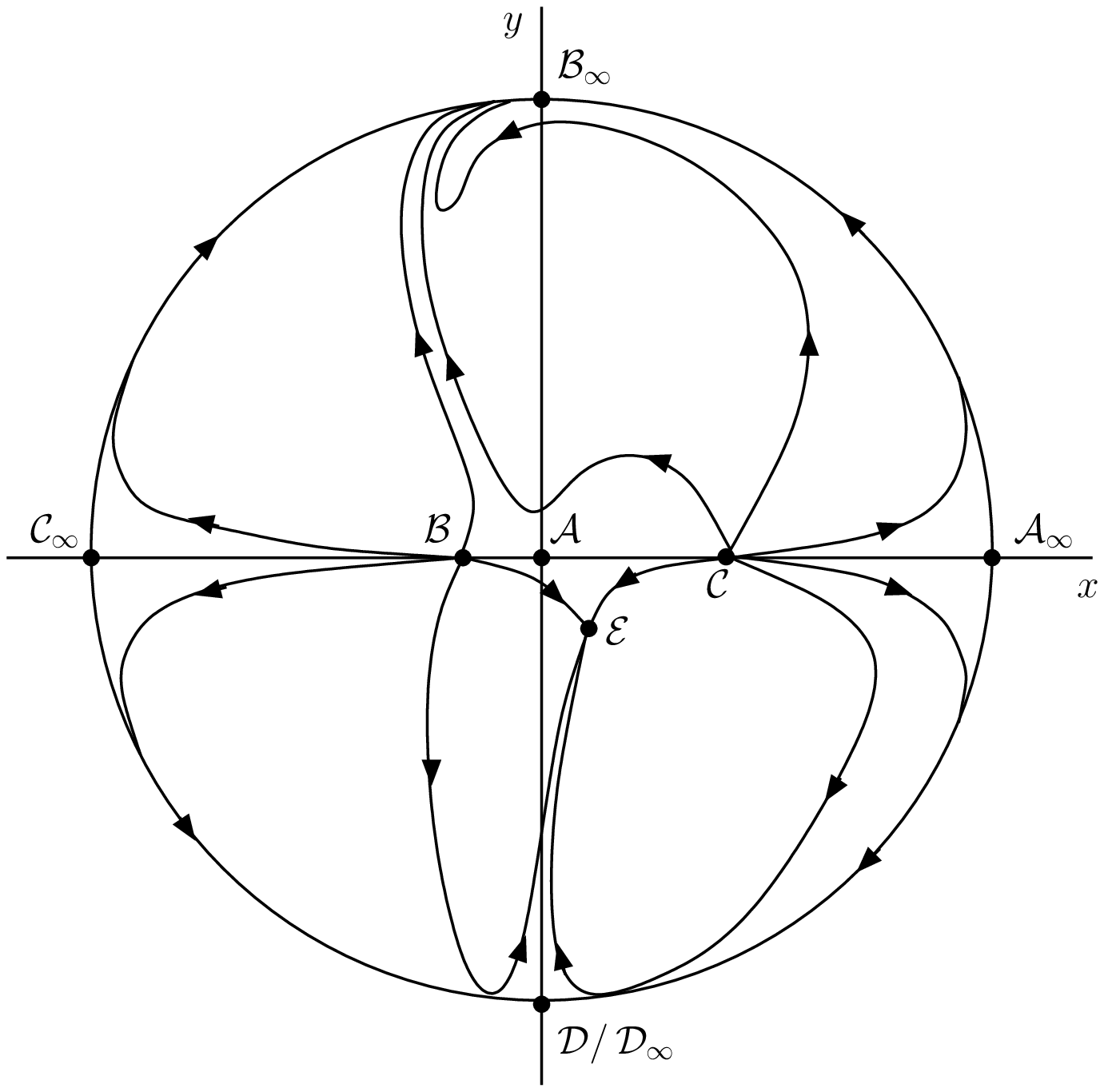} \caption{Global phase space for
$n=2$ with $\xi=-1$.}\label{fig2}
\end{center}
\end{figure}

\begin{figure}
\begin{center}
\includegraphics[clip=true,scale=0.50]{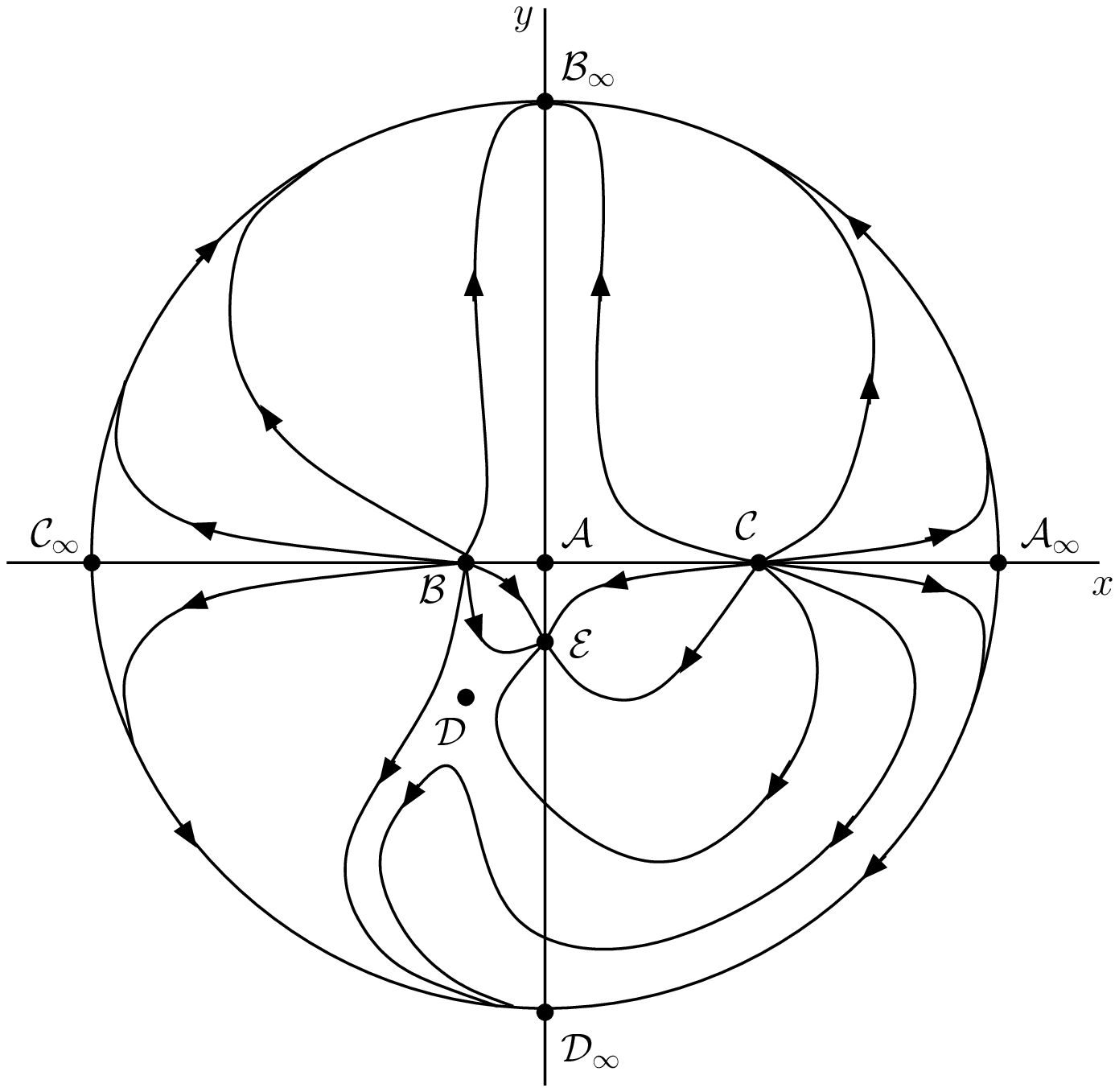} \caption{Global phase space for
$2<n<4+\sqrt{\frac{12\xi-1}{\xi}}$ (e.g $n=4$ and
$\xi=-1$).}\label{fig3}
\end{center}
\end{figure}

\begin{figure}
\begin{center}
\includegraphics[clip=true,scale=0.50]{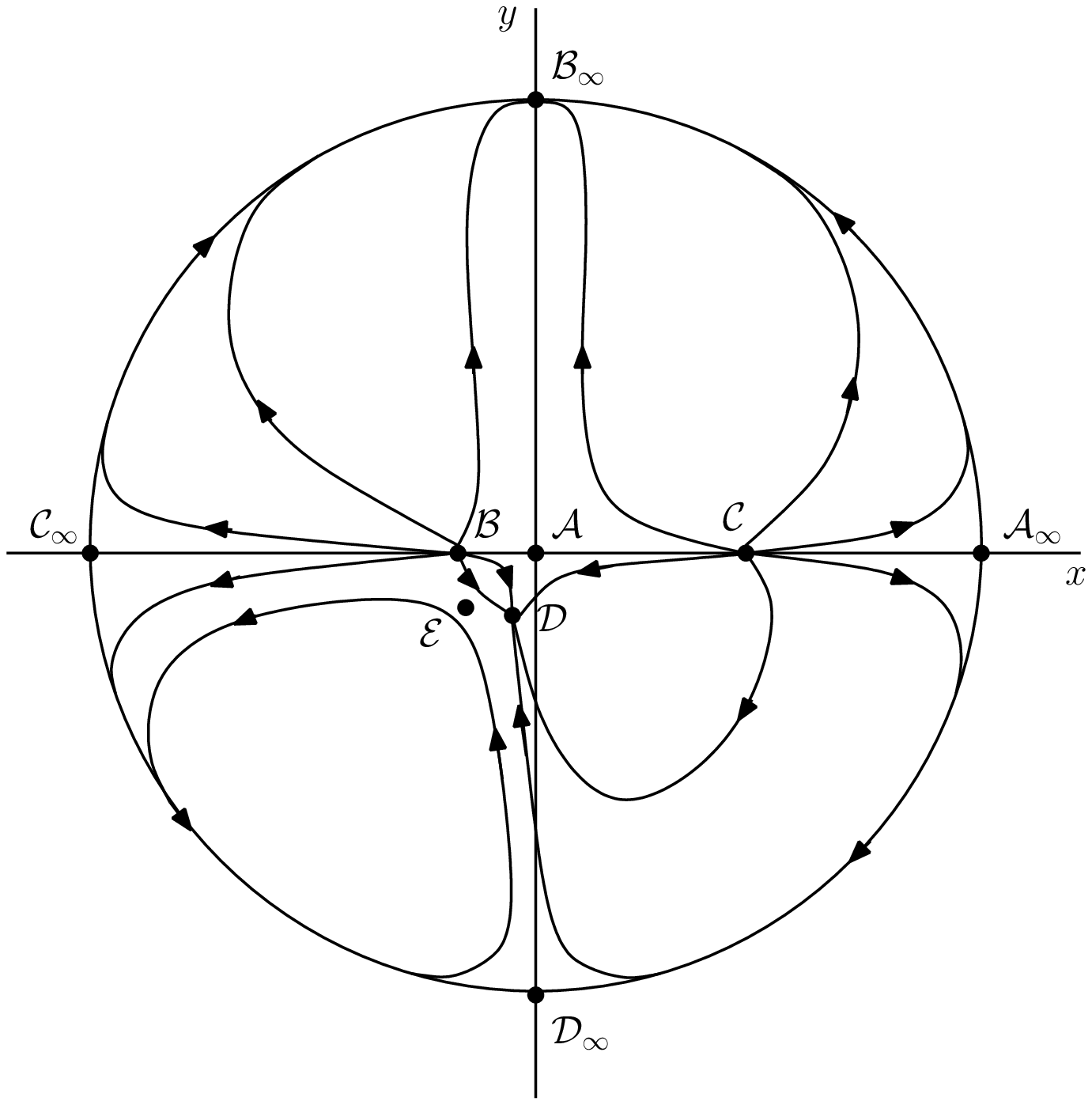} \caption{Global phase space for
$4+\sqrt{\frac{12\xi-1}{\xi}}<n<4+\sqrt{3}\sqrt{\frac{12\xi-1}{\xi}}$
(e.g. $n=8$ and $\xi=-1$).}\label{fig4}
\end{center}
\end{figure}

Since the phase space is two dimensional we can easily draw phase
space diagrams for the vacuum case. Here we will limit ourselves to
four examples representing the global phase space  \footnote{By
``global phase space"  we mean the projection of the $\theta> \pi
/2$ part of the Poincar\'{e} sphere on the plane that contains its
equator.} for four specific values of $(n,\xi)$ (see Figures
\ref{fig1}-\ref{fig4}) that includes the two cases in which the
theory admits s a GR attractor. 

\section{The matter case}

As we have seen in the system \rf{phi2:eqn_mat},  the presence of
matter in the dynamical system equations implies the introduction of
another variable: $z$. Since $\mu$ is defined to be positive, the
definition of the dynamical variables tells us that only negative
values of $z$ are compatible with attractive gravity. For this
reason we will restrict the following analysis only to the
physically relevant case $z\leqslant 0$.

As in the vacuum case, we can use the constraint
\rf{phi2:constraint} to eliminate one of the dynamical variables.
The system \rf{phi2:eqn_mat} then reduces to:
\begin{eqnarray}\label{phi2:eqn_mat2}
\hspace{-15 mm} x' = \frac{1}{12\xi-1}
\left[2(1-12\xi)x+4(1-12\xi)x^2-\left(2-\sfrac{1}{6\xi}\right)x^3
+6\xi (n-4)y\right. \nonumber \\
\left.-(1-6\xi(n-2))xy -\sfrac{1}{2}(1+3w-24\xi)xz+6\xi(1-3w)z \right], \nonumber \\
\hspace{-15
mm}y'=\frac{y}{12\xi-1}\left[2(12\xi-1)+(12\xi-1)(n-6)x+2\left(2-\sfrac{1}{6\xi}\right)x^2
\right. \nonumber \\
\left. -2(1-6\xi(n-2))y -(1+3w-24\xi)z  \right], \\
\hspace{-15 mm} z' = \frac{z}{12\xi-1}\left[(1+3w)(12\xi-1) -6(12\xi-1)x \right. \nonumber \\
\left. +2\left(2-\sfrac{1}{6\xi}\right)x^2-2(1-6\xi(n-2))y-
(1+3w-24\xi)z\right] \;. \nonumber
\end{eqnarray}
Since $y'=0$ and  $z'=0$ are zero for $y=0$ and $z=0$, the two
planes $y=0$ and $z=0$ corresponds to two invariant submanifolds.
The first plane represents classes of theories in which the
potential is zero, the second one classes of theories for which
$z=0$ and constitutes part of a vacuum invariant submanifold. The
structure of the total vacuum invariant submanifold can be derived
by writing the energy density in terms of the dynamical variables
\cite{Carloni05,Leach06a}:
\begin{equation}\label{DS:mu}
\mu \propto zy^{\frac{2}{n-2}}H^{\frac{2n}{n-2}}.
\end{equation}
It is clear that when $z=0$ and $y \neq 0$ the energy density is
zero. However when $y=0$ and $z \neq 0$ the behaviour of $\mu$
depends on the value of $n$: the energy density is zero when $n>2$,
but it is divergent when $n<2$. When both $y$ and $z$ are equal to
zero, we can conclude that $\mu=0$ only if $n>2$. For $n<2$
\rf{DS:mu} is divergent and $\mu$ can only be obtained by directly
solving the field equations.

\subsection{Finite analysis}

Setting $x'=0$, $y'=0$ and $z'=0$ we obtain eight fixed points (see
Table \ref{T:fixed_mat}). The first five ($\mathcal{A}$,
$\mathcal{B}$, $\mathcal{C}$, $\mathcal{D}$ and $\mathcal{E}$) sit
in the $z=0$ plane and have the same $(x,y)$ coordinates of the
corresponding vacuum fixed points. The coordinates of $\mathcal{F}$
and $\mathcal{G}$ are independent of $n$ and are finite for all
$\xi$; $\mathcal{H}$ is a finite fixed point for $n\neq0$ and merges
with the asymptotic fixed subspace $\mathcal{L}_\infty$ for $n=0$.

Merging occurs between the points $\mathcal{B}$, $\mathcal{C}$,
$\mathcal{D}$, $\mathcal{E}$, $\mathcal{G}$ and $\mathcal{H}$. The
first four points merge in the same way and for the same values of
the parameters given in the vacuum case. Point $\mathcal{H}$ merges
with  $\mathcal{E}$  for $n=\frac{1}{2}\left(7+3 w +\sqrt{\frac{\xi
\left(9 w ^2+66 w +73\right)-6 (w +1)}{\xi }}\right)$; and with
$\mathcal{G}$ when $n=\frac{3 (w +1) (8 \xi +w -1)}{4 \xi  (3 w
-1)}$.

All the fixed points but $\mathcal{A}$, $\mathcal{D}$, $\mathcal{F}$
are associated with flat solutions. $\mathcal{A}$ is associated with
negative curvature and the sign of the spatial curvature for point
$\mathcal{D}$ and $\mathcal{F}$ depends on the value of $n$ and
$\xi$.

The stability of the fixed points may be determined using the
Hartman-Grobman theorem  as in the vacuum case. The point
$\mathcal{A}$ is a saddle for every value of $w$. Point
$\mathcal{B}$ keeps the same stability found in the vacuum case for
$w=0$ and $1/3$, but it is always a saddle when $w=1$ . Instead, the
stability of points $\mathcal{C}$ and $\mathcal{D}$ is the same as
in the vacuum case for every value of $w$. Point $\mathcal{E}$
varies its stability  with $\xi,n,w$ and can behave like an
attractor, a repeller or a saddle as shown in
Table~\ref{T:stability_E_mat}. On the other hand, the fixed points
$\mathcal{F}$ and $\mathcal{G}$ are saddles for every value of the
parameters. Finally, $\mathcal{H}$ is a saddle or a saddle focus for
$w=0$ and $w=1/3$, while for $w=1$ it can also be a repeller or an
anti-spiral. Its stability is summarized in
Table~\ref{T:stability_H_mat}.

As in the vacuum case the coordinates of the fixed points can be
used to find exact solutions for the evolution of the scale factor.
From  \rf{Friedmann1:phi2} and \rf{KleinGordon:phi2} we have
\begin{eqnarray}
\fl\dot{H}+\frac{H^2}{\alpha_m}=0,\qquad \alpha_m
=\left[1-2x_i-\sfrac{1}{6\xi}x_i^2+\left(\sfrac{1+12\xi-6\xi
n}{1-12\xi}\right)y_i+\left(\sfrac{1+3w-24\xi}{2-24\xi}\right)z_i\right]^{-1}\label{phi2:mat_H}\;,\\
\fl\frac{\ddot{\phi}}{\phi}+\frac{\alpha_{m}^{2}\,\beta}{(t-t_{0})^{2}
}=0,\qquad \beta=3x_i+x_i^2+
\sfrac{6\xi(n-4)}{1-12\xi}y_i+\sfrac{6\xi(1-3w)}{1-12\xi}z_i\;,\label{phi2:mat_phi}
\end{eqnarray}
where $(x_i,y_i,z_i)$ represents the coordinates of the fixed
points. For $\alpha$ and $\beta$ different from zero, these
equations have solutions of the form \rf{phi2:vac_sol_gen} and
\rf{sol phi}.

The points $\mathcal{A}$, $\mathcal{B}$, $\mathcal{C}$,
$\mathcal{D}$ and $\mathcal{E}$ have the same solutions as in the
vacuum case, since for these points $z=0$. In particular the
convergence mechanism of scalar tensor gravity towards GR is
preserved when matter is present. Such a result was expected since
in \cite{Garcia-Bellido:1990jz} and \cite{Damour:1993id} this
phenomenon is described when matter is present. The point
$\mathcal{F}$ is associated to the Milne evolution and a constant
scalar field
\begin{equation}\label{phi2:mat_sol_F}
\alpha_{\mathcal{F}}=1\qquad\qquad \phi_{\mathcal{F}}=\phi_0\;,
\end{equation}
the value of the constant $a_0$ being influenced by the parameters
$\xi$ and $w$ for non-zero spatial curvature
\begin{equation}\label{phi2:mat_a0_F}
a_0^2=\frac{16k\xi}{1-16\xi+(2-3w)w}.
\end{equation}
However, a direct check of the equations reveals that this solution
is valid only for $w=-1/3$. Since in our model standard matter For point $\mathcal{G}$
we have
\begin{equation}\label{phi2:mat_sol_G}
\alpha_\mathcal{G}=\frac{2(8\xi-1+w)}{32\xi+3(w^2-1)},\qquad
m_\mathcal{G}= \frac{3 w ^2+48 \xi  w +16 \xi -3}{2 \left(3 w ^2+32
\xi-3\right)}\;.
\end{equation}
This solution represents a Friedmann like expansion with the exponent
depending on both $\xi$ and $w$.  For $0<w<1/3$ we have
$\displaystyle{\frac{1}{2}\leq\alpha_\mathcal{G}<\sfrac{2-16\xi}{3-32\xi}}$,
for $1/3\leq w<1$ we have
$\displaystyle{\frac{1}{3(1-8\xi)-4\sqrt{3\xi(12\xi-1)}}\leq\alpha_{\mathcal{G}}<\frac{1}{2}}$.
Direct comparison with the cosmological equations reveals that this
solution is physical only for $w>2/3$ and $w\neq1$.

Finally, point $\mathcal{H}$ with $n\neq 2$ is linked to a
solution that a first glance resembles a well known Friedmann GR one:
\begin{equation}\label{phi2:mat_sol_H}
\alpha_\mathcal{H}=\frac{2n}{3(n-2)(1+w)},\qquad
m_\mathcal{H}=\frac{n+2}{2 (n-2)}\;.
\end{equation}
This solution represents a Friedmann-like expansion for $n<0$ and
$\displaystyle{n>\frac{6 (w +1)}{3 w +1}}$, a contraction for
$0<n<2$ and power-law inflation when $\displaystyle{2<n<\frac{6 (w
+1)}{3 w +1}}$. When $n=2$ we obtain a solution of the form
\begin{equation}\label{phi2:mat_sol_H2}
a=a_0\, , \qquad \phi=\phi_0, \qquad \lambda=0,
\end{equation}
which corresponds to a static universe with cosmological constant.
Also this solution represent an effective ``GR state" for the scalar
tensor cosmology but in this case it is not a stable one. This means
that in our model of non-vacuum scalar tensor cosmology  no stable
non vacuum GR like solutions are allowed.

Using \rf{DS:mu} and the cosmological equations it is easy to
conclude that the points $\mathcal{F}$ and $\mathcal{G}$ both admit
vacuum solutions. This is not true for $\mathcal{H}$, for which the
energy density is
\begin{equation}\label{DS:mu_C}
\mu=\mu_0 t^{-\frac{2n}{n-2}},
\end{equation}
when $n\neq2$, where
\begin{eqnarray*}
\mu_0 &=& z_\mathcal{H}
\left(\frac{2y_\mathcal{H}}{\lambda}\right)^\frac{2}{n-2}
\left[\frac{4\xi n^2}{3(n-2)^2(1+w)^2} \right]^\frac{n}{n-2},
\end{eqnarray*}
and $y_\mathcal{H}$ and $z_\mathcal{H}$ are the $y$ and $z$
coordinates of $\mathcal{H}$ respectively. It is clear from the
expression above that this point does not represent a physical
solution for all values of $n$ and $\xi$ since $\mu_0$ and
therefore $\mu$ can, in principle, be negative. In order for us to
determine the values of $n$ and $\xi$ for which $\mathcal{H}$ is
physical, we have to solve the inequality $\mu_0>0$. We obtain
\begin{equation}
\fl \sfrac{1}{2}(7+3 w) -\sqrt{\sfrac{\xi  \left(9 w ^2+66 w
+73\right)-6 (w +1)}{4\xi }}<n<\sfrac{1}{2}(7+3 w) +\sqrt{\sfrac{\xi
\left(9 w ^2+66 w +73\right)-6 (w +1)}{4\xi }}\;.
\end{equation}
When $n=2$ the cosmological equations reveal that $\mathcal{H}$ is
instead associated with a vacuum ($\mu=0$) solution.

\begin{table}[tbp] \centering
\caption{The coordinates and the sign of the spatial curvature of
the non vacuum fixed points. }
\begin{tabular}{lll}
& & \\
\br Point &  Coordinates $(x,y,z)$ & $K$
\\ \hline
& & \\
$\mathcal{A}$ & $\left(0,\; 0,\; 0\right)$ &  $-1$  \\
$\mathcal{B}$ & $\left(2(-6\xi-\sqrt{3\xi(12\xi-1)}),\; 0,\;
0\right)$ &
$0$  \\
$\mathcal{C}$ & $\left(2(-6\xi+\sqrt{3\xi(12\xi-1)}),\; 0,\;
0\right)$ &
$0$  \\
$\mathcal{D}$ & $\left(-\frac{2}{n-2},\;
\frac{2-24\xi}{3\xi(n-2)^2},\; 0\right)$
&  $-\frac{1+[4+n(n-8)]\xi}{(n-2)^2\xi}$  \\
$\mathcal{E}$ & $\left(\frac{2(n-4)\xi}{1-2(n+2)\xi},\;
\frac{(12\xi-1)\left[3+(n-10)(n+2)\xi\right]
}{3(1-2(n+2)\xi)^2},\; 0\right)$ &  $0$  \\
$\mathcal{F}$ & $\left(-\frac{(1+3w)}{2},\; 0,\frac{(12\xi-1)(1+3w)}{12\xi}\right)$ &
$\frac{1-16\xi+(2-3w)w}{16\xi}$ \\
$\mathcal{G}$ & $\left(\frac{4\xi(1-3w)}{8\xi-1+w},\; 0,
\frac{(12\xi-1)\left[3(w-1)^2-16\xi(2-3w)\right]}{3(8\xi-1+w)^2} \right)$ &
$0$  \\
$\mathcal{H}$ & $\left(-\frac{3(1+w)}{n},
\frac{3-3w^2-4\xi \left[6(1+w)+n(1-3w)\right]}{4n^2\xi},\right.$ &
$0$ \\
& $\left. \frac{3(1+w)+2\xi\left[n^2-6(1+w)-n(7+3w)\right
]}{2n^2\xi}\right)$ & \\
& & \\ \br
\end{tabular}\label{T:fixed_mat}
\end{table}

\begin{table}[tbp] \centering
\caption{The exponent $\alpha$ of the scale factor solutions and the
energy density for the non vacuum case.}
\begin{tabular}{lll}
\br Point &  $\alpha$ & Matter density
\\\mr
$\mathcal{A}$ &  $1$ & $\mu=0$\\
$\mathcal{B}$ &
$\frac{1}{3(1-8\xi)-4\sqrt{3\xi(12\xi-1)}}$ & $\mu=0$ \\
$\mathcal{C}$ &  $\frac{1}{3(1-8\xi)+4\sqrt{3\xi(12\xi-1)}}$ & $\mu=0$ \\
$\mathcal{D}$ &  $1$, $\left(a_0^2=-\frac{k(n-2)^2\xi}{1+[4+n(n-8)]\xi}\right)$ & $\mu=0$\\
$\mathcal{E}$ &  $\left\{ \begin{tabular}{ll} $\frac{2(n+2)\xi-1}{(n-4)(n-2)\xi}$, & $n\neq 2,\;4$ \\
 $a=a_0 e^{C_0(t-t_0)}$, & $n= 2,\;4$ \end{tabular} \right.$ & $\mu=0$\\
$\mathcal{F}$ & $1$,
$\left(a_0^2=\frac{16k\xi}{1-16\xi+(2-3w)w}\right)$ & $\mu=0$ \\
$\mathcal{G}$ &  $\frac{2(8\xi-1+w)}{32\xi+3(w^2-1)}$ &
$\mu=0$  \\
$\mathcal{H}$ &
$\left\{ \begin{tabular}{ll} $\frac{2n}{3(n-2)(1+w)}$, & $n\neq 2$ \\
$a=a_0$, & $n= 2$\end{tabular} \right.$  & $\left\{
\begin{tabular}{l}
$\mu=\mu_0 t^{-\frac{2n}{n-2}}$ \\ $\mu=0$ \end{tabular} \right.$\\
\br
\end{tabular}\label{T:sol_mat}
\end{table}

\begin{table}[tbp] \centering
\caption{The parameter $m$ and the corresponding scalar field
solutions for the non vacuum case. The integration constants have
been calculated by direct substitution in the cosmological
equations.}
\begin{tabular}{lll}
\br Point &  $m$  & Solutions
\\ \mr
$\mathcal{A}$ & $\frac{1}{2}$ & $\phi=\phi_1$,\hspace{16mm} $\lambda=0$ \\
$\mathcal{B}$ & $\frac{3}{2 \left(24 \xi +4 \sqrt{3\xi(12 \xi
-1)}-3\right)}$
& $\phi=\phi_0 (t-t_0)^{1/2+m}$, \ $\lambda=0$\\
$\mathcal{C}$ &  $\frac{3}{2 \left(-24 \xi +4  \sqrt{3\xi(12 \xi
-1)} +3\right)}$
& $\phi=\phi_1 (t-t_0)^{1/2-m}$, \ $\lambda=0$\\
$\mathcal{D}$ & $\frac{n+2}{2 (n-2)}$  & $\left\{\begin{tabular}{ll}
$\phi=\phi_0^\mathcal{D} (t-t_0)^{1/2}$, & $n=-2$\\
$\phi=\phi_1^\mathcal{D} (t-t_0)^{1/2-m}$, & $n\neq -2,\;2$
\end{tabular} \right.$ \\
& & \\
$\mathcal{E}$ & $\frac{n+2}{2 (n-2)}$  & $\left\{\begin{tabular}{l}
$\phi=\phi_0^\mathcal{E} (t-t_0)^{1/2}$,\
\ \ \ $n=-2$\\
$\phi=\phi_1^\mathcal{E} (t-t_0)^{1/2-m}$,\ \ \ \  $n\neq -2,\;2$   \end{tabular} \right.$\\
& & \\
$\mathcal{F}$ &  $\frac{1}{2} (3 w +2)$ & $\phi=\phi_1$ ($w=-1/3$), \ \ \ $\lambda=0$ \\
$\mathcal{G}$ & $\frac{3 w ^2+48 \xi  w +16 \xi -3}{2 \left(3 w
^2+32 \xi-3\right)}$& 
$\lambda=0$, \quad $\xi=\frac{3(w-1)^{2}}{16(3w-2)} \rightarrow (w>2/3$, $w\neq 1$ ) \\
& & \\
$\mathcal{H}$ & $\frac{n+2}{2 (n-2)}$  & $\left\{\begin{tabular}{l}
$\phi=\phi_0^\mathcal{H} (t-t_0)^{1/2}$,\
\ \ \ $n=-2$\\
$\phi=\phi_1^\mathcal{H} (t-t_0)^{1/2-m}$,\ \ \ \  $n\neq -2,\;2$ \\
$\phi_0^\mathcal{H}=\pm2\left(\sfrac{24\lambda(1+w)^2}{3(1-w^2)-16\xi(1+3w)}\right)^{\frac{1}{4}}$,
\\
$\phi_1^\mathcal{H}=\left(\sfrac{2(3-4(n+6) +12(n-2)\xi
w-3w^2)}{3\lambda (n-2)^2(1+w)^2}\right)^{\frac{1}{n-2}}$
\end{tabular} \right.$\\\br
\end{tabular}\label{T:sol_mat_phi}
\end{table}

\begin{table}[tbp] \centering
\caption{The eigenvalues associated with the non vacuum fixed
points.}
\begin{tabular}{ll}
& \\
\br Point &  Eigenvalues \\ \hline
& \\
$\mathcal{A}$ &
$\left[-2,2,-1-3w\right]$  \\
$\mathcal{B}$ &
$\left[4(1-12\xi)-8\sqrt{3\xi(12\xi-1)},\; 6-2(n+2)\left(6\xi+\sqrt{3\xi(12\xi-1)}\right),\right.$ \\
& $\left. 3(1-8\xi)-\sqrt{3\xi(12\xi-1)}-3w \right]$  \\
$\mathcal{C}$ &
$\left[4(1-12\xi)+8\sqrt{3\xi(12\xi-1)},\; 6-2(n+2)\left(6\xi-\sqrt{3\xi(12\xi-1)}\right),\right.$ \\
& $\left. 3(1-8\xi)+\sqrt{3\xi(12\xi-1)}-3w \right]$  \\
$\mathcal{D}$ & $\left[\frac{(4-n)\xi-\sqrt{\xi
\{3(8-n)n\xi-4\}}}{(n-2)\xi},\;
\frac{(4-n)\xi+\sqrt{\xi\{3(8-n)n\xi-4\}}}{(n-2)\xi},\; \frac{6-n}{n-2}-3w \right]$  \\
$\mathcal{E}$ & $\left[\frac{3+(n-10)(n+2)\xi}{2(n+2)\xi-1},\;
\frac{2+2\{4+(n-8)n)\}\xi}{2(n+2)\xi-1},\; \frac{3+2\{-6+(n-7)n)\}\xi}{2(n+2)\xi-1}-3w \right]$  \\
$\mathcal{F}$ & $\left[\sfrac{1}{2}\left(4+(2-n)(1+3w)\right),\;
\frac{-2\xi(1-3w)-
\sqrt{2\xi\{18\xi(1+w)^2-(1-w)(1+3w)^2\}}}{4\xi},\right.$ \\
& $\left. \frac{-2\xi(1-3w)+ \sqrt{2\xi\{18\xi(1+w)^2-(1-w
)(1+3w)^2\}}}{4\xi} \right]$  \\
$\mathcal{G}$ & $\left[\frac{16\xi-(1-w)(1+3w)}{8\xi-1+w},\;
\frac{3(1-w)^2-16\xi(2-3w)}{2(8\xi-1+w)},\; \frac{3(w^2-1)+2\xi\{6+n+3(2-n)w\} }{8\xi-1+w} \right]$  \\
$\mathcal{H}$ & $\left[\frac{-4+(n+2)(1+3w)}{n},\;
\frac{3(w-1)(n+2)+12-\sqrt{S(n,\xi,\omega
)}}{4n},\; \frac{3(w-1)(n+2)+12+\sqrt{S(n,\xi,w)}}{4n} \right]$  \\
& \\
& {\small $S(n,\xi,w)=3\xi^{-1}( -1 + 12\xi)^{-1}\left\{
-36(w-1)(1+w)^2 +
3\xi \left[ 4(1+w)^2(12w-37)\right. \right.$} \\
&  {\small $\left. + n^2\left(7+(2-9w)w \right) + 4n(1+w)
\left(w(19+6w)-17\right)\right]  + 4\xi^2\left[ 324(1+w)^2 \right.$}
\\
& {\small $\left. \left. + 8n^3(-1+3w)+ 12n(1+w)(29+3w) +n^2\left(17
-21w(10+3w)\right)\right]\right\}$} \\
 & \\\br
\end{tabular}\label{T:eigenvalues_mat}
\end{table}

\begin{table}[tbp] \centering
\caption{Stability of the fixed point $\mathcal{E}$ for the matter
case. The parameters are $N_{\pm}=4\pm
\sqrt{3}\sqrt{(12\xi-1)/\xi}$, $P_{\pm}^{w}=\frac{1}{2}(7+3w)\pm
\sqrt{\frac{\xi(9w^2+66w+73)-6(1+w)}{4\xi}}$ and $Q_{\pm}=4\pm
\sqrt{(12\xi-1)/\xi}$.}
\begin{tabular}{lcccc}
\br  & {\small $n< \frac{1-4\xi}{2\xi}$} & {\small
$\frac{1-4\xi}{2\xi}< n<N_- $}
& {\small $N_{-}<n<P_{-} $} & {\small $P_{-} <n<Q_{-} $} \\
\mr {\small $w=0, 1/3$} & {\small attractor} & {\small
repeller} & {\small saddle} & {\small saddle}\\
{\small $w=1$} & {\small attractor} & {\small
repeller} & {\small saddle} & {\small saddle}\\
\br & {\small $Q_{-}< n< Q_{+}$} & {\small $Q_{+}<n< P_{+} $} &
{\small $P_{+}<n<N_{+} $} & {\small $N_{+}<n<P^1_{+} $} \\ \mr
{\small $w=0, 1/3$} & {\small attractor} & {\small
saddle} & {\small saddle} & {\small repeller}\\
{\small $w=1$} & {\small attractor} & {\small
saddle} & {\small saddle} & {\small saddle}\\
\br & {\small $n> P^1_{+}$} &
&  &  \\
\mr{\small $w=0, 1/3$} & {\small repeller} &  &  & \\
{\small $w=1$} & {\small repeller} &  &  &  \\
\br
\end{tabular}\label{T:stability_E_mat}
\end{table}

\begin{table}[tbp] \centering
\caption{Stability of the fixed point $\mathcal{H}$ of the matter
case. The parameters are $P_{\pm}=\frac{1}{2}(7+3w)\pm
\sqrt{\frac{\xi(9w^2+66w+73)-6(1+w)}{4\xi}}$ and $S_{i}$, which are
the real roots of the polynomial $S(n,\xi,w)$ given in
Table~\ref{T:eigenvalues_mat}, for the given value of $w$. We use
the term `anti-spiral' to denote a pure repulsive spiral and `saddle
focus' to denote an unstable point with two complex eigenvalues.  }
\begin{tabular}{lcccc}
\br {\small $w=0$} & {\small $n\leq S_1 $} & {\small $S_1<n<S_2 $}
& {\small $S_2\leq n \leq S_3 $} & {\small $n>S_3$} \\
\mr  & {\small saddle} & {\small
saddle focus} & {\small saddle} & {\small saddle focus}\\
& {\small ($n\neq \frac{3(1-8\xi)}{4\xi}$)} &  & {\small ($n\neq 0,6,P_\pm$)}& \\
\br {\small $w=1/3$} & {\small $n\leq S_1 $} & {\small $S_1<n<S_2 $}
& {\small $n>S_2$} &  \\
\mr & {\small saddle focus} & {\small
saddle} & {\small saddle focus} & \\
& & {\small( $n\neq 0,4,P_\pm$)} & & \\
\br {\small $w=1$} & {\small $n\leq S_1 $} & {\small $S_1\leq n<P_+
$} &
{\small $P_+< n \leq S_2 $} & {\small $S_2<n<S_3$} \\
\hline &
{\small saddle focus} & {\small
saddle} & {\small repeller} & {\small anti-spiral}\\
& & {\small ($n\neq 0,3,P_-$)} & & \\
\br
 & {\small $S_3\leq n <6 $} & {\small $n>6 $} &  &  \\ \mr & {\small
repeller} & {\small
saddle} &  & \\
\br
\end{tabular}\label{T:stability_H_mat}
\end{table}

As in the vacuum case we can write the deceleration parameter $q$
in terms of the dynamical variables via
\begin{equation}
\fl
q=-\frac{\dot{H}}{H^{2}}-1=2x_i+\sfrac{1}{6\xi}x_i^2-\left(\sfrac{1+12\xi-6\xi
n}{1-12\xi}\right)y_i-\left(\sfrac{1+3w-24\xi}{2-24\xi}\right)z_i.
\end{equation}
The surface defined above divides the phase space in two volumes
representing the region in which the expansion is accelerated and
the region in which it is decelerated. The fixed points
$\mathcal{A}$, $\mathcal{B}$, $\mathcal{C}$, $\mathcal{D}$ and
$\mathcal{E}$ behave like their vacuum counterparts, since they all
satisfy $z=0$. Point $\mathcal{F}$ lies on this surface consistently
with the fact that it corresponds  to Milne evolution. The point
$\mathcal{G}$ always lies in the region in which the expansion is
decelerating. The behaviour of $\mathcal{H}$ depends on the
barotropic factor $w$ and the parameter $n$: it lies in the
accelerated evolution region for $w=0$ when $n<0$ and $n>6$, for
$w=1/3$ when $n<0$ and $n>4$ and for $w=1$ when $n<0$ and $n>3$. Otherwise it lies in the decelerated evolution region.

\subsection{Asymptotic analysis}

We next study the asymptotic behaviour of the system
\rf{phi2:eqn_mat2} using the Poincar\'{e} projection. The
compactification of the phase space can be achieved by transforming
to spherical coordinates
\begin{equation}
x=\bar{r}\sin \theta \cos \psi, \ \ \ \ y=\bar{r}\sin \theta \sin
\psi, \ \ \ \ z=\bar{r} \cos \theta,
\end{equation}
where $\bar{r}=\frac{\sqrt{r}}{1-r}$ and $\bar{r} \in [0,\infty) $,
$\theta \in [0,\pi]$ and  $\psi \in [0, 2\pi]$. In the limit $r
\rightarrow 1$ ($\bar{r} \rightarrow \infty$), equations
\rf{phi2:eqn_mat2} become
\begin{eqnarray}
r' = \frac{\cos^2 \psi\; \sin^2 \theta}{24\xi}\left[(\cos 2 \psi -3)
\sin ^2\theta-4 \cos ^2\theta  \right], \label{phi2:inf_r_mat} \\
\theta' = -\frac{\cos \theta \; \sin^3 \theta \; \cos^4 \psi }{6\xi(1-r)^{2}},  \label{phi2:inf_th_mat} \\
\psi ' = \frac{\sin^2 \theta\; \cos^3 \psi\; \sin
\psi}{6\xi(1-r)^{2}}. \label{phi2:inf_phi_mat}
\end{eqnarray}
As in the vacuum case the radial equation does not contain the
radial coordinate, so that the fixed points can be obtained using
only the angular equations. Setting $\psi '=0$ and $\theta '=0$, we
obtain four fixed points and a fixed subspace which are listed in
Table~\ref{Table_phi2:mat_asymp}.

The points $\mathcal{A}_\infty$ and $\mathcal{B}_\infty$ lie on the
plane $z=0$ and therefore their solutions are the same as the vacuum
points $\mathcal{A}_\infty$ and $\mathcal{C}_\infty$, respectively.
Points $\mathcal{C}_\infty$ and $\mathcal{D}_\infty$ represent the
poles of the Poincar\'{e} sphere and it is easy to prove that they
are linked to the same solutions at $\mathcal{A}_\infty$ and
$\mathcal{B}_\infty$. The solution for the scalar field instead
reads
\begin{equation}
 \phi=c_1 e^
{c_0(t-t_0)\sqrt{\sfrac{1+3w-24\xi}{2-24\xi}}}+c_2e^{
c_0(t-t_0)\sqrt{\sfrac{1+3w-24\xi}{2-24\xi}}}.
\end{equation}

In addition to the fixed points above, we found two  fixed subspaces
$\mathcal{L}_1$ and $\mathcal{L}_2$, that contains all the points
with $\psi=\pi/2$ and $\psi=3\pi/2$, respectively. These subspaces
corresponds to $x\to 0$, $y/z\to \pm \tan\theta_0$, respectively.
The equations for $y'$ and $z'$ in this limit reduce to
\begin{eqnarray}
y'&=&\frac{y^2}{1-12 \xi
   } \left[2(1-6 \xi(n-2)  )\pm (1+3 w-24 \xi ) \cot \theta_0\right]\;, \\
z' &=& \frac{z^2}{1-12 \xi} \left[(1+3 w-24 \xi )\pm 2(1-6 \xi(n-2)
) \tan \theta_0 \right]\;,
\end{eqnarray}
which admit the solutions
\begin{eqnarray}
y &=& \frac{12 \xi-1 }{(\tau-\tau_{\infty}) \left[2(1-6 \xi(n-2))\pm
(1+3 w-24 \xi ) \cot \theta_0\right]\;,}\label{ycerchio}\\
z &=& \frac{12 \xi -1}{(\tau-\tau_{\infty}) \left[(1+3 w-24 \xi )\pm
2(1-6 \xi(n-2)) \tan \theta_0\right]\;.}\label{zcerchio}
\end{eqnarray}
In the limit above, \rf{phi2:mat_H} becomes
\begin{equation}\label{hcerchio}
    \dot{H}=-\left[\left(\sfrac{1+12\xi-6\xi
n}{1-12\xi}\right)y+\left(\sfrac{1+3w-24\xi}{2-24\xi}\right)z\right]H^2\;,
\end{equation}
which implies
\begin{eqnarray}\label{solcerchio}
 |\tau-\tau_{\infty}|=\left[C_1 \pm C_2(t-t_0)\right]^{2}.
 \end{eqnarray}
The solution for the scalar field in the fixed subspaces
$\mathcal{L}_1$ and $\mathcal{L}_2$, can be found in similar way. In
the limit above, using \rf{ycerchio}, \rf{zcerchio} and
\rf{solcerchio} in \rf{phi2:mat_phi} we find
\begin{equation}\label{eqficerchi}
\frac{\ddot{\phi}}{\phi}+A c_0^2=0\;,
\end{equation}
where
\begin{eqnarray}\label{solcerchio1}
A= \frac{6\xi(1-3w)\cos \theta_0 \pm 12\xi(n-4)\sin \theta_0}{2(1+3
w-24 \xi )\cos \theta_0 \pm 4(1-6 \xi(n-2)  ) \sin \theta_0}\;.
\end{eqnarray}
We find the following solutions
\begin{equation}\label{SolFiCerchio}
\fl \phi= \left\{\begin{array}{ll} c_1\cos
(\sqrt{|A|}c_0(t-t_0))+c_2\sin (\sqrt{|A|}c_0(t-t_0)),
  & \mbox{if}\; A<0,\\
  &\\
c_1 e^ {\sqrt{A}c_0(t-t_0)}+c_2e^{\sqrt{A}c_0(t-t_0)}, & \mbox{if}\;
A>0. \end{array}\right.
\end{equation}

The stability of these asymptotic fixed points can be deduced by
analyzing the stability with respect to the angular coordinates, and
from the sign of $r'$. The points $\mathcal{A}_\infty$ and
$\mathcal{B}_\infty$ are stable nodes for all values of $n$ and
$\xi$. Points $\mathcal{C}_\infty$ and $\mathcal{D}_\infty$
instead are always saddles.

All the other points including the fixed subspaces $\mathcal{L}_1$
and $\mathcal{L}_2$, have both eigenvalues equal to zero. In order
to derive the stability of these points we have to analyze the
effect of the non-linear contributions  have on dynamical equations.
This can be done by Taylor developing the R.H.S 
of the dynamical equations around the fixed point up to the first non-zero
order and then  directly the resulting system.

For the points $\mathcal{C}_\infty$ and $\mathcal{D}_\infty$ we
obtain the solutions:
\begin{eqnarray}
\psi_{C_\infty} = c_2, \quad  \quad \quad&& \theta_{C_\infty}  =
-\frac{6 \xi }{\tau \sin \left(\psi _0\right) \cos ^3\left(\psi
   _0\right)+6 \xi  c_1}, \\
\psi_{D_\infty} = c_2, \quad \quad \quad && \theta_{D_\infty} = \pi
-\frac{6 \xi }{\tau \sin \left(\psi _0\right) \cos ^3\left(\psi
_0\right)+6 \xi  c_1}.
\end{eqnarray}
This result tells us that when we choose initial conditions
around these points, the evolution of the universe will follow an
orbit with constant $\psi$, with  $\tau$ increasing
\footnote{\label{nota}This behaviour is reversed in case of a
contracting cosmology because in this case $\tau$ effectively
changes sign when $a$ is decreasing.} and $\theta$ approaching  0 or
$\pi$. If we also consider the behaviour of the radial equation we
conclude that these points behave like saddles.

For the fixed subspaces $\mathcal{L}_1$ and $\mathcal{L}_2$, we find
\begin{eqnarray}
 \fl \psi (\tau) &=& \frac{\sqrt[3]{-2\xi }}{\sqrt[3]{\tau \cos \theta_{0}
   \sin ^3\theta_{0}+96 \xi  c_1}}\pm\frac{\pi }{2}\;, \\
 \fl \theta (\tau) &=&\theta_{0}-\frac{\tan \theta_{0}}{2}+ c_2 \left[\tau\left(\sin4
   \theta_{0}-2  \sin 2 \theta_{0} \right)-768 \xi  c_1)\right]^{\frac{2}{3}\csc ^2\theta_{0}}\;.
\end{eqnarray}
It is clear that, for $\tau$ increasing, the solution for $\psi$
approaches $\pi/2$ for $\mathcal{L}_1$ and $3\pi/2$ for
$\mathcal{L}_2$, while $\theta$ increases. The radial behaviour is
very complicated and depends critically on the value of the coordinate
$\theta_0$, $w$, $n$ and $\xi$ (we will not show this dependence here). We may
however conclude that this subspace is a saddle or an attractor
depending on the value of $\theta_{0}$.

\begin{table}[tbp] \centering
\caption{Coordinates, behaviours and stability of the asymptotic
fixed points in the non vacuum case. The value of the parameter $A$
is given in \rf{solcerchio1}.}
\begin{tabular}{llll}
\br Point & $(\theta,\psi)$ & Scale factor & Stability
\\ \mr
$\mathcal{A}_\infty$ & $(\sfrac{\pi}{2},0)$
&$|\tau-\tau_\infty|=\left[C_1 \pm C_2(t-t_0)\right]^{2}$ &
attractor
 \\
$\mathcal{B}_\infty$ & $(\sfrac{\pi}{2},\pi)$ &
$|\tau-\tau_\infty|=\left[C_1 \pm
C_2(t-t_0)\right]^{2}$ & attractor   \\
$\mathcal{C}_\infty$ & $(0,0)$ & $|\tau-\tau_\infty|=\left[C_1 \pm
C_2(t-t_0)\right]^{2}$ & saddle  \\
$\mathcal{D}_\infty$ & $(\pi,\pi)$ &
$|\tau-\tau_\infty|=\left[C_1 \pm C_2(t-t_0)\right]^{2}$ & saddle \\
\mr $\mathcal{L}_1$ & $(\theta_0,\sfrac{\pi}{2})$ &
$|\tau-\tau_\infty|=\left[C_1 \pm C_2(t-t_0)\right]^{2}$ &
\\
$\mathcal{L}_2$ & $(\theta_0,\sfrac{3\pi}{2})$ &
$|\tau-\tau_\infty|=\left[C_1 \pm C_2(t-t_0)\right]^{2}$ &
\\
\br
\end{tabular}\label{Table_phi2:mat_asymp}
\end{table}
The phase space in the matter case is 3-dimensional and therefore
cannot be visualized  as easily as its vacuum counterpart. For this
reasons  we will not give here any sketch of the phase space and we
refer the reader to the next section for an analysis of the results
derived above.

\section{Discussion and conclusions}
In this paper, the dynamics of STG FLRW cosmological models has been
studied using a phase space analysis. We have considered a
generic non--minimally coupled theory of gravity where the coupling
and the potential are powers of the scalar field, both in a vacuum and
in presence of a perfect fluid. The set of parameters characterizing
the cosmological models are $\{\xi,\lambda,n,w\}$, i.e. the coupling
constant, a constant parameter, the power of the self-interacting
potential and the barotropic index of the perfect matter fluid,
respectively. The phase-space is 3-dimensional in absence of matter
while it is 4-dimensional in presence of matter, but in both
cases the FLRW Hamiltonian constraint allows one to reduce their
dimensionality. Our investigation considered the existence and
local stability of critical points (finite analysis) and the
asymptotic analysis via the Poincar\'{e} projection.

We identified five finite fixed points in the vacuum case and eight in the
matter case. In the vacuum case, there are four asymptotic stability
points corresponding to the four intersections of the axes with the
unitary Poincar\'{e} circle. In the matter case, we have to consider
a unitary 3-sphere and in additin to the vacuum asymptotic points we
find two more points and two ``fixed subspaces". The stability of
fixed points strictly depends on the values of the above parameters
and in particular on $\xi$ and $n$, i.e. the coupling and the power
of self--interacting potential.

In the vacuum case, the two dimensional phase space is divided in
two halves by the invariant submanifold associated with $V(\phi)=0$.
Of the nine fixed points we found, five of them
($\mathcal{A},\mathcal{B},\mathcal{C},\mathcal{A}_{\infty},\mathcal{C}_{\infty}$)
are permanently on this invariant submanifold, two of them
$\mathcal{B}_{\infty},\mathcal{C}_{\infty}$ do not change their
position and the other two ``move" in the phase plane depending on
the values of the parameters. In particular, $\mathcal{D}$ is
characterized by $q=0$ and its position is on the curve described by
the equation given in (\ref{q}) and $\mathcal{E}$ is associated with
a flat spatial geometry and always lies on the curve described by
the constraint (\ref{phi2:constraint_vac}) with $K=0$. Even if
matter is not present the scalar field is able to induce an
expanding Friedmann-like evolution. However, the value of $\alpha$
in (\ref{phi2:vac_sol_B}) and (\ref{phi2:vac_sol_C}) reveals that at
these fixed points the scale factor cannot grow faster than
$t^{1/2}$. In addition, these solutions are unstable for every value
of the parameters. The point $\mathcal{E}$ admits the widest
spectrum of behaviours. Depending on the values of $n$ and $\xi$ its
solution can represent an inflationary phase, a Friedmann-like phase
or a contraction. A comparison with the stability analysis reveals
that $\mathcal{E}$ is an attractor only when it corresponds to
contracting solutions or power law inflation. This has two
consequences: (i) in the scalar field dominated regime our model of
scalar tensor cosmology admits an inflationary phase as an attractor
even if this attractor is not global and not unique; (ii) since
there is no value of the parameters for which the $\mathcal{E}$  is
a saddle, this model does not admit a transient inflationary phase
and cannot solve the graceful exit problem.

The case $n=4$ is particularly interesting because, for this value of
$n$,  $\mathcal{E}$ is associated with a GR ($\phi=const.$) de
Sitter solution and is an attractor. This fact has two main
consequences: (i) it shows that with a non--minimally coupled scalar
field with quartic potential, the early time cosmology evolves
towards an inflationary phase which is indistinguishable from a GR
one; and (ii) it gives us an independent confirmation of the idea
that a scalar tensor theory of gravity can evolve towards GR.  The
difference is that when a potential is present, the realization of
such mechanism is strictly related to the form of the potential.

In the matter case, we found three new finite fixed points which are
physical only for specific values of the parameters. In particular,
$\mathcal{F}$ is never physical because it satisfies the
cosmological equations only for a negative $w$. Point $\mathcal{G}$
represents a physical vacuum solution only for $w>2/3$ and $w\neq
1$. For these values of $w$, this point is associated with an
expansion whose rate cannot be higher than the radiation dominated
GR-FLRW solution. Point $\mathcal{H}$ is associated to a non-vacuum
solution which resembles a well known Friedmann-GR solution. As $n$
varies, this solution can represent power law inflation, a
decelerated Friedmann solution and a contraction. For $n=2$, this
point is linked  to a GR-like state, but since it is unstable for
every value of the parameters, we conclude that there is no way for
this class of scalar tensor cosmologies to approach a stable
non-vacuum GR state. In fact, a quick look to the table above
reveals that there is no value of the parameters for which any of
the finite fixed points are stable. This means that the only
non-vacuum attractors for this class of theories are the asymptotic
fixed points and their associated Lema\^{\i}tre solution. Therefore
we can conclude that the class of cosmologies we treated are doomed
either to approach an effectively vacuum state (probably
corresponding to thermal death) that corresponds to one of the
vacuum attractors or to recollapses towards a Big Crunch.

The most interesting orbits are the ones that ``travel" between
point $\mathcal{E}$ and $\mathcal{H}$. This is because they could
represent cosmic histories in which a Friedmann-like cosmology
enters naturally in a phase of accelerated expansion or cosmic
histories in which an unstable inflationary phase is followed by a
second inflationary phase. The first type of orbits is interesting
because they can in principle help solve  the incompatibility
between the evolution towards a Dark Energy era and the formation of
large scale structure. The second ones are interesting because they potentially
unify ``dark" scalar fields and the inflaton within a single scheme.

Since the phase space is three dimensional it is not easy to check
if such orbits actually exist without the use of numerical
techniques. However, our results allow one to give some necessary
condition for these orbits to exist and to rule out some of them.
For example, it is easy to see that there is no value of the
parameters for which $\mathcal{E}$ can represent an unstable
Friedmann solution and $\mathcal{H}$ an inflationary phase. On the
other hand, for $4<n<Q_+$ and $w=0,1/3$ we have that $\mathcal{H}$
corresponds to an unstable decelerated expansion, $\mathcal{E}$
correspond to a stable power law inflation and these points are not
separated by any invariant submanifold. This means that, in
principle cosmic histories exist for which a transient Friedmann
evolution approaches to a power law inflationary phase in a natural
way.

If $4<n<6(1+w)/(3w+1)$ and $w=0,1/3$ another interesting set of
cosmic histories is possible in which we have two inflationary
phases, a first one which is unstable associated with $\mathcal{H}$
and a second one which is  stable associated with $\mathcal{E}$.
Although only a detailed analysis of these cosmic histories can
reveal if this last group of cosmic histories also include the
deceleration phase necessary for the realization of standard
cosmology, this scenario is interesting because it shows that in STG
the non--minimally coupled scalar field can act as {\em both} the
inflaton and dark energy. Such a behavior has been also found in a
class of higher order gravity models \cite{shosho} and definitely
deserves a more careful investigation. Such a study, will
be the topic of a a forthcoming paper.

As final comment, it is worth stressing the connection between the
model presented above and the String-Dilaton action. Using the
transformation (\ref{Trasf:String}), which is only a
reparametrization for the scalar field, we can pass from the action
\rf{action:F(phi)} to \rf{action:string}. This means that all the
exact solutions for the scale factor that we have derived together
with their stability are also solutions for the String-Dilaton
action. Therefore, our analysis also allows us to give details of
the dynamics  of low-energy string cosmology which could be of great
interest in the quest for finding observational constraints for this
theory.

\ack

This research was supported by the National Research Foundation
(South Africa) and the Italian {\it Ministero Degli Affari Esteri-DG
per la Promozione e Cooperazione Culturale} under the joint Italy/
South Africa Science and Technology agreement. The authors wish to
thank A Starobinsky and G Venturi for their useful comments.


\section*{References}

\end{document}